\documentstyle[12pt]{article}
\let\ssection=\section
\renewcommand{\section}{\setcounter{equation}{0}\ssection}

\newcommand\mathC{\mkern1mu\raise2.2pt\hbox{$\scriptscriptstyle|$%%@
}
		{\mkern-7mu\rm C}}				% The complex  numbers 
\newcommand{\mathR}{{\rm I\! R}}         % The real numbers

\begin{document}

\begin{center}
{\large The End of Time?}
\end{center} 
\begin{center}
Jeremy Butterfield, All Souls College, Oxford OX1 4AL,
England
\end{center}
\begin{abstract} I discuss Julian Barbour's Machian theories of %%@
dynamics, and his proposal that a Machian perspective enables one %%@
to solve the problem of time in quantum geometrodynamics (by %%@
saying that there is no time!). I concentrate on his recent book %%@
{\em  The End of Time} (1999). A shortened version will appear in %%@
{\em British Journal for Philosophy of Science}.  
\end{abstract}
\noindent{\bf 1}\/ {\em Introduction}\\
{\bf 2}\/ {\em Machian themes in classical physics}\\
\indent{\bf 2.1}\/{\em The status quo}\\
\indent\indent {\bf 2.1.1} {\em Orthodoxy}\\
\indent\indent {\bf 2.1.2} {\em The consistency problem}\\
\indent{\bf 2.2 }\/{\em Machianism}\\
\indent\indent {\bf 2.2.1} {\em The temporal metric as %%@
emergent}\\
\indent\indent {\bf 2.2.2} {\em Machian theories}\\
\indent\indent {\bf 2.2.3} {\em Assessing intrinsic dynamics}\\
{\bf 3}\/ {\em The end of time?}\\
\indent{\bf 3.1}\/{\em Time unreal? The classical case}\\
\indent\indent {\bf 3.1.1} {\em Detenserism and presentism}\\
\indent\indent {\bf 3.1.2} {\em Spontaneity}\\
\indent\indent {\bf 3.1.3} {\em Barbour's vision: time %%@
capsules}\\
\indent{\bf 3.2}\/{\em Evidence from quantum physics?}\\ 
\indent\indent {\bf 3.2.1} {\em Suggestions from Bell}\\
\indent\indent {\bf 3.2.2} {\em Solving the problem of time?}

\section{Introduction}
Barbour is a physicist and historian of physics, whose research %%@
has
for some thirty years focussed on Machian themes in the %%@
foundations of
dynamics. There have been three main lines of work: in classical
physics, quantum physics, and history of physics---as follows. He %%@
has
developed novel Machian theories of classical dynamics (both of
point-particles and fields), and given a Machian analysis of the
structure of general relativity; (some of this work was done in
collaboration with Bertotti). As regards quantum physics, he has
developed a Machian perspective on quantum geometrodynamics; this %%@
is
an approach to the quantization of general relativity, which was
pioneered by Wheeler and DeWitt, and had its hey-day from about %%@
1965
to 1985. More specifically, Barbour proposes that a Machian %%@
perspective
enables one to solve an outstanding conceptual problem %%@
confronting
quantum geometrodynamics, the so-called `problem of time'.  In %%@
short,
his proposal is that there is no time! (Hence the title of this
book.) As regards history, he has uncovered the tangled tale of %%@
the
reception (and often misinterpretation) of Mach's ideas in %%@
twentieth
century physics (including general relativity); and also written %%@
a
two-volume history of the theory of motion, stretching from the %%@
ancient Greeks
to the twentieth century.

Barbour has now written what is in effect an intellectual
autobiography, in the form of a book of popular science ([1999]: %%@
page references are to this book). It covers all
three of his lines of work. It of course emphasises the first %%@
two,
especially the second (the speculative proposals about quantum
gravity). The work on classical physics is discussed in Parts 2 %%@
and 3
of the book; and the quantum speculations in Parts 1, 4 and 5. %%@
But
Barbour also weaves in a considerable amount of historical %%@
material.

 Although it is a popular book, there are three reasons, of %%@
increasing
importance, to review it in this Journal.  The first relates to %%@
the
style of the book: it is not a hack popularization. Like all good %%@
popular
science, it makes a real attempt to expound the details, both of
established theories and speculative proposals; rather than just
stating the main idea---or worse, just gesturing at it with a %%@
metaphor
which is liable to be as misleading as it is helpful. (The %%@
obvious
comparison here is with e.g. Penrose's {\em The Emperor's New %%@
Mind},
as against Hawking's {\em A Brief History of Time}; or as regards
magazines, with e.g. {\em Scientific American}, as against {\em %%@
New
Scientist}.) The book contains plenty of detailed
exposition---exposition often enlivened by metaphors that are %%@
helpful
as well as vivid. (Some readers will also enjoy the anecdotes of %%@
Barbour's
various intellectual struggles, journeyings and collaborations.)
Barbour also moderates his passionate advocacy of his ideas,
especially his quantum speculations, with occasional reminders %%@
that
they are controversial.

The second reason is that, although {\em cognoscenti} will know
Barbour's previous work on Machian themes in classical physics %%@
(both
relativistic and non-relativistic), it is worth having an %%@
accessible
exposition in a single book. For in some respects, this work goes
against prevailing opinion in the philosophy of space and time; %%@
and
perhaps for this reason, it is still not very well known. I shall %%@
devote Section 2 to this material. To set the scene, I
shall first describe prevailing opinion, and a problem it faces. %%@
Then in discussing Barbour's Machian theories, I will emphasise %%@
how Barbour takes
space rather than spacetime as fundamental, and how this is in %%@
tension
with relativity.\footnote{Fortunately, Barbour's work on %%@
classical
physics is now becoming better known: Belot ([1999], Sections %%@
6-7; [2000], Section 4)
discusses it; and Pooley ([2001]), to which I am much indebted, %%@
is a
full analysis, both technically and philosophically. An early
philosophical discussion by Barbour himself is in this journal
([1982]). I should add that as to Barbour's historical work, the %%@
main
source is the first volume ([1989]) of his two-volume history; it
covers the history of dynamics up to Newton. The second volume of
course includes a history of Machian ideas in the twentieth %%@
century,
but is not yet published; in the meantime, Barbour ([1999a]) is %%@
an
accessible summary of that history.}

The third reason relates to Barbour's denial of time. %%@
Philosophers of
physics, and indeed metaphysicians, are bound to want to know %%@
what
this denial amounts to. Fortunately, I can present the main ideas %%@
in terms of
familiar metaphysical categories, without recourse to quantum %%@
theory,
let alone quantum gravity. I shall do so in Section 3.1; we shall %%@
see
that it is a curious, but coherent, position which combines %%@
aspects of
modal realism {\em \`{a} la} David Lewis and presentism {\em %%@
\`{a} la}
Arthur Prior.  Then finally, in Section 3.2, I shall discuss how
Barbour argues for his denial of time from certain claims about %%@
the interpretation of 
quantum theory, and about quantum gravity.\footnote{For a brief %%@
discussion in this journal of these claims, cf. Brown ([1996]). }

But beware: although my strategy of postponing quantum theory and %%@
quantum gravity, with all their obscurities, until Section 3.2 %%@
makes for an exposition more accessible to philosophers, it also %%@
carries a price. Namely, it emphasizes those aspects of Barbour's %%@
denial of time which can be explained in terms of classical %%@
physics (i.e. roughly, in terms of instantaneous configurations %%@
of matter), in particular Barbour's idea of a `time capsule'; and %%@
it downplays a technical quantum-theoretic aspect, which Barbour %%@
(private communication) sees as prior to, and more important %%@
than, time capsules. So I should at the outset summarize this %%@
aspect, `the price', and say why I think my strategy is %%@
justified.

Barbour believes that his Machian analysis of general relativity %%@
gives the best understanding of (and justification for) the two %%@
equations that sum up the theory in the form in which it is most %%@
easily quantized. (The equations are called the momentum and the %%@
Hamiltonian constraint equations; the form of the theory is %%@
called the Hamiltonian, or canonical, form.) Since quantizing %%@
general relativity (in this form) by an otherwise successful %%@
method leads to a static i.e. time-independent quantum state, %%@
Barbour concludes that we must accept such a state and somehow %%@
reconcile it with the appearance of time and change. He takes %%@
this to be his main conclusion: and time capsules to be his %%@
admittedly conjectural suggestion for how to make the %%@
reconciliation.

Nevertheless, there is good reason for me to emphasize time %%@
capsules, at the expense of the arguments leading to a static %%@
quantum state. For as we shall see in Section 3.2, most of these %%@
arguments have been well known in the physics literature for many %%@
years, and some are even prominent in the growing philosophical %%@
literature about quantum gravity. Besides, the literature %%@
contains several (mostly very technical) strategies for %%@
reconciling a static quantum state with the appearance of time %%@
and change. Barbour's time capsules proposal is but one of these, %%@
with the advantage that it can be explained non-technically: so %%@
in a review of his work, I have of course chosen to focus on it, %%@
ignoring the others.

I shall finish this Introduction with my three main criticisms of %%@
the
book as a popularization. First, the book gives the misleading
impression that Barbour's various views are closely connected one %%@
with
another. In fact, Barbour's views are by no means a
package-deal. In particular, the straightforward and %%@
craftsmanlike
work in classical physics and in history of physics can be %%@
`bought';
while the denial of time, and the speculations about quantum %%@
theory
and quantum gravity, are left on the shelf.

The other two criticisms both concern `the end of time'; the %%@
first
from a metaphysical viewpoint, the second from a physical one.  %%@
First:
Barbour often (not just in the `prospectus', Part 1, but also in %%@
Parts
4 and 5) states his denial of time in a way that philosophers %%@
will
immediately interpret as just denying temporal becoming. This is
misleading: his view {\em is} different from the familiar
tenseless (`B-theory') view of time---as we will see in Section %%@
3.

Second: As I mentioned above, the quantum gravity programme on %%@
which
Barbour focusses, quantum geometrodynamics, has been superseded.
Agreed, it has a distinguished descendant, the loop quantum %%@
gravity programme, which is one of the two main current %%@
programmes; (the other being the superstrings
programme, which is utterly different, not just technically, but %%@
also
in its motivations and framework). But the central notion of a %%@
configuration is very different in loop quantum gravity than in %%@
quantum geometrodynamics; and Barbour has yet to tell us how his %%@
proposals, e.g. about time capsules, would carry over to it. %%@
Besides, even on its own terms,
quantum geometrodynamics is very problematic. In particular, its %%@
main
equation, the Wheeler-DeWitt equation, is not mathematically
well-defined, and its `derivation' is questionable. Though
Barbour briefly mentions the rival programmes, and the quantum
geometrodynamics programme's internal difficulties (at pp. 38-39, %%@
166,
192, 351), the bulk of the book is devoted to presenting his %%@
Machian
vision of exactly this programme. So a reader who is a newcomer %%@
to the
subject will inevitably get the impression that geometrodynamics %%@
is
still `one of the runners' in the race of quantum gravity %%@
research.
Agreed, that race has no clear running-track or rules: it is %%@
rather like
orienteering in a blizzard---without a map! So indeed, 
quantum geometrodynamics just might `come from the back of the %%@
field to win the race'. But newcomers be warned: it seems %%@
unlikely.\footnote{Agreed, these considerations of physics do not %%@
nullify
the metaphysical interest of articulating and assessing Barbour's
denial of time---cf. Section 3.1. And his denial of
time just might help with the conceptual problems faced by the %%@
rival programmes.}

These three criticisms aside, the book is, overall, very good %%@
popular
science: we can happily go along with Barbour when he announces %%@
that
he has `tried to write primarily for the general reader ... but %%@
shall
be more than happy if my colleagues look over my shoulder' (p. %%@
7).

\section{Machian themes in classical physics}
 I shall first summarize the prevailing opinion in the philosophy %%@
of
space and time, as three claims; and describe how they face a
consistency problem. This problem is not insoluble, nor %%@
unrecognized;
but it is substantive (Section 2.1). This will set the scene for %%@
discussing Barbour's
own work (Section 2.2).

\subsection{The status quo}
\subsubsection{Orthodoxy}
In contemporary philosophy of space and time, the prevailing %%@
opinion
is that the development of physics from the mid-nineteenth %%@
century
(especially the rise of field theories culminating in general
relativity) was the death-knell of the relationist tradition, %%@
stemming
from Leibniz to Mach, of conceiving space and time as systems of
spatiotemporal relations between bodies, rather than entities in %%@
their
own right. More precisely, there are three prevalent claims
which need to be articulated here. I shall list them in what I %%@
believe
to be the order of increasing controversy!\footnote{Anyway, we %%@
will
see in Section 2.2, that Barbour agrees
with the first two claims, and maybe with the third. I should %%@
also explain that my phrase `prevailing opinion' refers to the
last 30 years. Before that, Reichenbach and others had maintained %%@
that
relativity's abolition of absolute space and time vindicated
Leibniz's and Mach's relationism against Newton's absolutism.  As %%@
we
will see, Barbour could, and I think would, agree that the {\em
detail} of Reichenbach's position is wrong, as argued by the %%@
`young
Turks' of the 1960s and 1970s---authors such as Stein, Earman and %%@
Friedman; (cf.  e.g. Earman [1989], p. 6f.). But Barbour would %%@
add that
Reichenbach misinterprets Mach, as egregiously as 
he does Newton!}  These claims concern, respectively: (1) field %%@
vs.
matter, (2) whether metrical structure is `reducible' to matter, %%@
and (3)
the status of spacetime points.

(1) The first claim concerns the rise of field theory since 1850.
Thus, on the one hand, physics revealed some traditional
characteristics of bodies---such as impenetrability and
continuity---to be only apparent. And on the other hand, the
electromagnetic field was discovered to possess mechanical %%@
properties
like momentum, angular momentum, energy and (after the advent of %%@
special
relativity) mass-energy. Besides, matter itself was eventually
modelled, with outstanding success, as a field on spacetime; %%@
namely in
quantum field theories. Thus Leibniz's and Mach's `bodies' had %%@
become
diaphanous and omnipresent fields.

(2) Second, the various familiar theories---both classical and
quantum, relativistic and non-relativistic---postulate a metrical
structure of spacetime that seems not to be
`reducible to spatiotemporal relations of bodies' (in various %%@
precise
senses of that phrase).  Two well known (probably the best known) %%@
ways
of making this claim precise concern: (a) absolute rotation, and %%@
(b)
dynamical metrics. 

For (a), the idea of the argument goes back to Newton's globes
thought-experiment. The two cases---one in which the mutually
stationary globes rotate at some constant velocity about their %%@
common
centre of gravity, and the other with no rotation---are intended %%@
to
show that the metric is irreducible; the idea is that the two %%@
cases
exhibit the same spatiotemporal relations of bodies but different
metrical facts. But one needs to be careful about `metrical %%@
facts'.
For in any theory, such as Newtonian mechanics or special %%@
relativity,
in which the metric structure is the `same in all models' (and so %%@
is
not dynamical, i.e.  not correlated with the distribution of %%@
matter),
there will be a sense of `metrical facts' in which these facts %%@
cannot
differ between two models---and so are trivially determined
by, i.e. supervenient upon, spatiotemporal relations between %%@
bodies. (Compare the idea in modal
metaphysics that a proposition $M$ true in all of a certain class %%@
of
worlds trivially supervenes on the truth of any proposition $S$ %%@
true
at some of the worlds: for any two worlds in the class that make %%@
$S$
true, also make $M$ true.)

There is an obvious strategy for otherwise interpreting `metrical
facts' in such a way that pairs of cases like the two globes %%@
nevertheless show
some kind of irreducibility of the metric.  The idea is that such %%@
a
pair of cases shows that not all physical possibilities are
distinguished by a full description of each of them in terms of %%@
the
masses, relative distances and relative velocities of bodies; so %%@
that
there are physically real properties or relations not determined %%@
by
these. What these are will differ from one proposal to another. %%@
But
irreducibility will follow, provided the `metrical facts' include %%@
some
of the facts involving such undetermined properties or relations. %%@
Here
the obvious candidate is facts about how the bodies are
related to the affine structure of spacetime---in short, about %%@
whether their
worldlines are geodesics.\footnote{However, perhaps one does not %%@
have to
express such facts only in terms of such `absolutist' postulates. %%@
Some
authors assume that the theory concerned includes gravity, so %%@
that the
non-rotating globes would fall towards each other, while the %%@
rotating
pair can be assumed to rotate at just the speed that compensates
gravity and gives a stable orbit. On this approach, one can take %%@
the
moral to be that the distribution of bodies over a period of time %%@
is
not reducible to the masses, relative distances and relative
velocities of the bodies at an initial instant---even in a
deterministic theory like Newtonian gravitational theory. Thanks %%@
to Oliver Pooley and Simon Saunders for this point.}

As to (b): For a theory with a dynamical metric such as general %%@
relativity
(or any theory that `geometrizes gravity' in the sense of coding
gravitational force as an aspect of the metric), the above %%@
problem of
`metrical facts' being `the same in all models', and so trivially
determined by the matter distribution, does not arise. Nor is the
correlation (or even causal relation) between the metric and the
matter distribution, as coded in the field equations, a sign of
reduction or even determination (i.e. supervenience). First, we %%@
cannot in general specify the matter distribution without the %%@
metric. Second,
general relativity admits pairs of solutions which agree on their
matter distributions but differ metrically. For general %%@
relativity,
the most-cited cases are the various vacuum (i.e.  matter fields %%@
zero)
solutions. But there are other examples, e.g.  the Schwarzschild
and Kerr metrics---representing respectively, a non-rotating and %%@
a
rotating mass in an otherwise empty universe.\footnote{Agreed, %%@
this is also an idealized, indeed vacuum example, in that for %%@
both solutions, the central mass is at a singularity which is not %%@
{\em in} the spacetime; and I for one do not know whether there %%@
are such pairs of solutions with extended matter in the %%@
spacetime. In any case, even if such
pairs show that the metric is not determined by, i.e. %%@
supervenient
upon, matter, there are ``Machian effects'', such as %%@
inertia-dragging,
in general relativity.  Some were found as early as 1918; indeed,
Einstein found similar effects in the course of his struggle %%@
towards
general relativity.}

(3) Third, the formalisms of field theories, both classical and
quantum, suggest that their basic objects are just spacetime %%@
points.
For in these formalisms, everything apart from these %%@
points---matter,
whether conceived as point-particles or as fields, and even the
metrical structure of spacetime---gets represented as %%@
mathematical
structures defined on points. This suggests the doctrine nowadays
called `substantivalism': that spacetime points are genuine %%@
objects---indeed are the basic objects of these theories, in that %%@
everything else is to
be construed as properties of points, or relations between them, %%@
or 
higher-order properties and relations defined over them.

Agreed, this doctrine is controversial even among those with no
sympathy for relationism. There are two well-known good (though
disputable!) reasons for denying it. (a): The first reason is %%@
`the
hole argument'. It is natural to formulate these theories, %%@
especially
those with a dynamical metric (like general relativity), in such %%@
a way
that their only `fixed structure' is the local topological and
differential structure of a manifold. Intuitively, this means %%@
that any
map preserving this structure (i.e. any diffeomorphism) preserves %%@
the
content of the theory. Such a theory is called `diffeomorphism
invariant'; and diffeomorphism invariance is often taken to %%@
indicate
that spacetime points are indeed not objects: their appearing to %%@
be is
an artefact of how we formulate the theory.  (b): The second %%@
reason is
a general philosophical point about objects or ontology.  Namely: %%@
we
should be wary of taking as the basic objects of our ontology
(according to some theory) those items that are postulated as the
initial elements in a mathematical presentation of the theory. %%@
For it
might be just a happenstance of our formulation of the theory %%@
that
these objects `come first': a happenstance avoided by another
formulation that can be agreed, or at least argued, to be better.

\subsubsection{The consistency problem}
These three claims lead to what I will call `the consistency %%@
problem'. This
is not a straightforward matter of the claims entailing the %%@
problem.
Rather, these claims, and the presentations of spacetime theories %%@
that
are now typical, both in physics textbooks and the philosophy of
physics literature, make us think of the metrical structure of
spacetime in a certain way---roughly, as `on all fours' with the
matter fields.\footnote{It is tempting to formulate the %%@
conception as
treating the metric as intrinsic to spacetime. But this is not %%@
quite
right. For although the notion of a property being intrinsic to %%@
an
object, and correspondingly of a relation being intrinsic to the
collection of its relata, is hard to analyse (witness recent %%@
effort in
analytic metaphysics, e.g. Langton and Lewis ([1998])), people's
judgments of what is intrinsic tend to agree. And it seems that a
metric that was wholly reducible to matter fields might yet be %%@
called
`intrinsic to spacetime'; especially if, as usual, the theory
postulates a manifold of spacetime points which then function as %%@
the
relata of the intrinsic metric. In Section 2.2, we will see that
Barbour works with  such theories; so his Machianism seems %%@
content
with the metric being intrinsic to spacetime in this sense.} And %%@
it is
this conception that leads to the consistency problem. As I said %%@
at
the start of the Section, I do not claim that this problem is
insoluble: rather it is a lacuna that needs to be filled. Nor is %%@
the
problem unrecognized: as we will see, Einstein himself recognized %%@
it
(and dubbed it a problem of `consistency'). Furthermore, we can %%@
nowadays state in broad terms how the problem can be solved (the %%@
lacuna filled). But the problem is worth stressing, since it is %%@
rarely
addressed---and it leads into Barbour's Machian proposals.

I shall present the problem by first
describing the lacuna in typical modern presentations of %%@
spacetime
theories. In such presentations, one postulates first a spacetime %%@
manifold;
then, some metric (and affine) structure on it, typically as %%@
metric
fields; then also some matter fields; these fields, both metric %%@
and
matter, obey differential equations; etc. So far, so good. %%@
Indeed, so
far, so pure mathematical. The connection with observation is %%@
then
made, for metric fields, by postulating that (ideal) rods and %%@
clocks
measure the metric, i.e. yield its values as their %%@
pointer-readings.
Or maybe the connection is made more generally than {\em via} the
proverbial rods and clocks---by postulating that the behaviour
(perhaps ideal) of matter fields gives us access, in principle, %%@
to the
metric fields. But---here is the rub---there is, typically, %%@
little or
no discussion of how the rods and clocks are constructed and how %%@
they
`do their stuff'---display pointer-readings that are veridical %%@
about
spacetime's metric. (Similarly, for more general accounts of how
matter gives access to metric.) If one does not notice this
lacuna\footnote{As is tempting! For example, think of the almost
mesmerising power of the purely diagrammatic explanation of
time-dilation and length-contraction in special relativity, using %%@
just
the hyperbolae that are the locus of points of constant Minkowski
interval from the origin---and the postulate that rods and clocks
measure this interval!}, one naturally ignores the question of %%@
how
indeed they do their stuff; and so falls into thinking of the %%@
metric
as simply `read off' by rods and clocks.

But of course it is a question that cannot be ignored. In any %%@
branch
of physics, one has a right to expect a theory of how the %%@
instrument
works (very often calling on other branches of physics, on which %%@
the
instrument's functioning depends). In this regard rods and clocks %%@
(or whatever
else, such as radar signals, is postulated as measuring the %%@
metric)
are no different from other instruments.

An account that filled this lacuna would proceed: from (i) the
postulation of fundamental metric and matter fields on spacetime; %%@
to
(ii) a theoretical description of how (at least idealized) rods %%@
and
clocks (or whatever else) behave, powerful and accurate enough to
secure that: (iii) such rods and clocks (or other matter) would %%@
indeed
do their stuff---their pointer-readings report the metric. There %%@
are
four remarks to make about such an account. Broadly speaking, %%@
they are positive: there are good prospects for such an account, %%@
though the details would of course be very complicated. (Or %%@
rather, there are good prospects if we set aside the usual deep %%@
controversies about how the macroscopic world emerges from the %%@
quantum one.) 

\begin{itemize}
\item The way that (iii) returns to (i) is in no way a suspicious %%@
circularity. An account
of how things are should be compatible with the explanation how %%@
we
come to know or believe that very account; and it is even better %%@
if
the account coheres with, or even is itself a part of, that
explanation.
\item As I mentioned, Einstein himself emphasised the need for %%@
such an account, and called the lacuna a problem of %%@
`consistency'. In his {\em Autobiographical Notes}, he writes in %%@
connection with special relativity:
\begin{quote}
One is struck [by the fact] that the theory introduces two kinds %%@
of
physical things, i.e. (1) measuring rods and clocks, (2) all %%@
other
things, e.g. the electro-magnetic field, the material point, etc. %%@
This
in a certain sense is inconsistent; strictly speaking, measuring %%@
rods
and clocks would have to be represented as solutions of the basic
equations (objects consisting of moving atomic configurations), %%@
not,
as it were, as theoretically self-sufficient entities. However, %%@
the
procedure justifies itself because it was clear from the very
beginning that the postulates of the theory are not strong enough %%@
to
deduce from them sufficiently complete equations for physical %%@
events
sufficiently free from arbitariness, in order to base upon such a
foundation a theory of measuring rods and clocks ... it was %%@
better to
permit such inconsistency---with the obligation, however, of
eliminating it at a later stage of the theory. But one must not
legalize the mentioned sin so far as to imagine that intervals %%@
are
physical entities of a special type, intrinsically different from
other physical variables (``reducing physics to geometry'', %%@
etc.).
([1949], pp. 59-60.)
\end{quote}
\item So the consistency problem is recognized. Indeed, some %%@
expositions of relativity,  less imbued
with the geometric style of presentation, sketch such an account %%@
of
rods and clocks: trying to discharge, in Einstein's phrase, the
obligation of eliminating the inconsistency. I say `sketch' %%@
because it
is very hard to describe, in terms of one's fundamental physical
theory, most types of clock (and most types of any other %%@
instrument):
Einstein notes this for special relativity, but the point is of %%@
course
equally valid for other theories, such as continuum mechanics in
Newtonian spacetime. As a result, the expositions that sketch %%@
such an
account often focus on some artificially simple models; e.g.  the
light-clock, or (as a model of length-contraction) the distortion %%@
in a
classical atom of the electron's orbit, due to the atom's motion; %%@
cf.
Bell [1976].\footnote{Some of the philosophical literature also
recognizes the consistency problem (though without using that %%@
label).
For example, Brown ([1993], [1997]) and Brown and Pooley ([2001])
discuss how it bears on the foundations of special and general
relativity.}

\item Besides, we have good reason to think we can go beyond such %%@
simple models, and in particular accommodate the quantum nature %%@
of matter. To take one example, let me sketch a quantum %%@
description of length-contraction for a moving rod. (Many thanks %%@
to David Wallace for this sketch.) I suppose that we start with %%@
some quantum field theory of electrons, the electromagnetic %%@
field, and a charged field of much higher charge and mass than %%@
the electron field (to simulate atomic nuclei of some particular %%@
element). This is a reasonable place to start, because: (i) we %%@
certainly need to go beyond classical
physics since it cannot describe solid bodies 
 other than phenomenologically (e.g. stipulating {\em a
priori} that rods are rigid); (ii) on the other hand, to require %%@
that we start with the
Standard Model and work our way up through quark-quark couplings
etc. would be unduly harsh---since after all, we don't believe it %%@
provides the final theory of matter. 

We then need to establish that there are certain states of this %%@
multi-field system describing a non-relativistic regime of %%@
particles coupled
by effective Coulomb forces, with respect to a 3+1 split given by %%@
some
Lorentz frame; and that in this regime, there are states %%@
corresponding to rigid non-relativistic matter, say a crystal. %%@
Suppose we succeed in this. Of course this is conceptually and %%@
technically difficult, given the interpretative
problems of quantum field theory, and the complexities of solid %%@
state physics. But given that there are such states, we can argue %%@
that a rod made of this crystalline matter will exhibit length %%@
contraction; as follows. 

If the rod is accelerated gently enough, the internal vibrations %%@
(phonons) due to the acceleration will be small and will rapidly %%@
thermalise (heating the crystal up slightly until the heat is %%@
lost to the environment), and the crystal
will  enter the state ${\hat B}\mid\psi\rangle$; where ${\hat B}$ %%@
is the
Galilean boost operator for the final velocity attained, and %%@
$\mid\psi\rangle$ is the original state of the rod i.e. the state %%@
of a rigid crystal with length (say) $l$ given by the
metric in our initial frame. The fact that sufficiently
small Lorentz boosts can be approximated by Galilean boosts then %%@
implies that we can iterate boosts to move $\mid\psi\rangle$ into %%@
the state ${\hat B_L}\mid\psi\rangle$,
where state ${\hat B_L}$ is any Lorentz boost. Since the %%@
underlying quantum-field-theoretic dynamical laws
are Lorentz-covariant, it follows that ${\hat %%@
B_L}\mid\psi\rangle$ has length $l$
as given by the metric in the boosted frame.  So we have %%@
established that the crystal accurately
measures the metric in its rest frame, i.e. length
contraction. (Or more accurately, we would have established this %%@
if our sketch were filled in!).   

\end{itemize}

To sum up: though one faces a consistency problem if one %%@
postulates
a metric `on all fours' with matter fields, our current theories %%@
of matter seem able to solve the problem, at least in outline.

\subsection{Machianism}
In this subsection, I will first make two general points about %%@
Barbour's variety of Machianism, and discuss his attitude to %%@
Section 2.1.1's three
claims (1) to (3) (Section 2.2.1).  This will lead to 
 some details of his Machian theories (Section
2.2.2). I will end with some evaluative comments about these
proposals, especially the central ingredient of the Machian
theories---the notion of an instantaneous configuration (Section
2.2.3). 

\subsubsection{The temporal metric as emergent}
One naturally expects that {\em ceteris paribus}, the consistency %%@
problem will be easier to solve, the weaker the metrical %%@
structure of spacetime that one postulates `on all fours' with %%@
matter fields: for there will be less structure that, for %%@
consistency, instruments' pointer-readings have to `reveal'.
On the other hand, one needs to postulate enough structure to be %%@
able to write down a dynamical theory: one cannot expect all such %%@
structure to be reducible to matter (cf. (2) of Section 2.1.1). %%@
So the question arises whether there are interesting dynamical %%@
theories that postulate less metrical structure than the familiar %%@
theories. Barbour's work shows that the answer is `Yes'. 

There are two immediate points to make about this `Yes'. First, %%@
in the theories Barbour has developed so
far, the gain mostly concerns the temporal aspect of the %%@
spacetime metric: though the temporal metric `emerges' from the %%@
rest of physics, the spatial metric is still %%@
postulated.\footnote{This is not to say that Barbour faces a %%@
consistency problem {\em only} for how rods measure spatial %%@
geometry, and not at all for how clocks measure time. But as we %%@
shall see, the problem of how a clock measures time will be %%@
easier for Barbour, in the sense that his Machian theory in a %%@
sense explains why different clocks ``march in step''. }

Second, a point about the notion of `emergence'. Usually, this %%@
notion is vague: and philosophers think of
making it more precise in terms of various definitions of %%@
reduction,
or some weaker analogue, such as supervenience. It will turn
out that in the context of Barbour's views, two more precise %%@
versions
play a role, one for the classical context (this Section), and %%@
one for
the context of quantum gravity (Section 3).  In the classical %%@
context,
Barbour's notion is logically strong, indeed surprisingly so. For
Barbour provides examples of theories in which a temporal (though %%@
not
spatial) metric is emergent in the strong sense of being fully
definable from the rest of the physical theory. So this is %%@
emergence
in as strong a sense of reduction as you might want. On the other
hand, in the context of quantum gravity, Barbour's notion is %%@
logically
weak (though again the detail is surprising). For his denial that
there is time leads to his saying that our illusion to the
contrary---that there is time---is a partial and misleading
`appearance' of the timeless underlying reality.

I turn to discussing Barbour's attitude to Section 2.1.1's %%@
claims, (1)
to (3), of `prevailing opinion'; in particular to justifying my
statement (in footnote 4) that he agrees to (1) and (2), and
maybe also (3). Since claim (2), about the irreducibility of %%@
metric to
matter, concerns `the various familiar theories', it is already %%@
clear
that he can accept it while nevertheless seeking theories in %%@
which there is some kind of reducibility.
The kind of theory he seeks emerges more clearly from his replies %%@
to
(1) and (3).

As to the claim (1), Barbour accepts that physics since 1850 has
`abolished body'. But he takes this to mean just that we should %%@
state
Machian proposals in terms of the modern conception of
matter, namely as matter fields. So although Barbour presents %%@
Machian
theories of point-particles (pp. 71-86, 115-120; and similarly
in his technical articles), that is intended as a piece of %%@
pedagogy
and/or heuristics---a path I will follow in Section 2.2.2.  As he
stresses, the main physical ideas of these theories carry over to
field theories.

This of course prompts the question: how exactly does Barbour %%@
conceive
field theories?  In particular, does he accept the existence of
spacetime points, as advocated by claim (3)?  It will be clearest %%@
to
break the answer to these questions into two parts. The first %%@
part
concerns Barbour's different attitudes to spacetime, and to %%@
space:
this part relates closely to the physical details of his %%@
proposals,
and will engender several  comments. The second part concerns
Barbour's relationist understanding of space; it is more
philosophical, and will crop up again later---at this stage, it %%@
can be
dealt with briefly.

The first part of the answer is, in short, that Barbour seems %%@
happy to
accept the existence of {\em spatial} (but not spacetime) points, %%@
and
to postulate that these points form a 3-dimensional manifold with
metric (spatial geometry). For he discusses field theories
(relativistic and non-relativistic) in terms of the evolution %%@
through
time of instantaneous 3-dimensional configurations, i.e.  states %%@
of a
3-dimensional spatial manifold. But one needs to be careful; (and %%@
here
lies the second part of the answer). Since Barbour seeks a %%@
relationist
understanding of space, he wants to treat metrical relations in a
field theory, not in terms of a metric field tensor on a %%@
3-manifold of
spatial points, but as very similar to inter-particle distances %%@
in
theories of point-particles. This kind of treatment will become
clearer in Section 2.2.2.2. Here it suffices to say two things. %%@
(i)
The main idea is that, just as a Machian theory of $N$ %%@
point-particles
will treat metrical relations as essentially a matter of %%@
$N(N-1)/2$
inter-particle distances, a Machian field theory will postulate
point-like parts of a matter field and then treat metrical %%@
relations
in terms of the continuous infinity of all the pairwise distances
between these point-like parts. (ii) As we shall see in a moment %%@
(and
again in Sections 2.2.2.2 and 2.2.3), this sort of Machian %%@
treatment of field
theories has difficulties with general relativity.

The first part of the answer---that Barbour postulates space, and
spatial geometry, but is leary of spacetime---also needs to be
clarified in other ways. Indeed, it holds good for his discussion %%@
of
both classical non-relativistic field theories, and quantum %%@
theories.
By his lights, spacetime notions have no central role in either %%@
of
these types of theory. But his reasons are rather different for %%@
the
two types (as I hinted above, in discussing strong and weak
emergence).  For classical non-relativistic field theories, the
temporal structure (and so spacetime structure) is fully defined, %%@
in
Barbour's Machian theories, by the rest of the theory: but ({\em
modulo} the attempt to understand space relationally) these %%@
theories
postulate a spatial manifold and metric {\em ab initio}; (cf. %%@
Section
2.2.2). On the other hand, for quantum theories, time (and so
spacetime) emerges only as an approximately valid notion within a
timeless quantum physics of matter, which again postulates {\em %%@
ab
initio} a spatial manifold with a metric; (cf. Section 3.2).

But Barbour also qualifies this acceptance of space but not %%@
spacetime in two other ways. First, he admits (pp. 180-181) that %%@
for classical
relativistic field theories, and especially for general %%@
relativity,
spacetime notions {\em are} central.  Though he discusses these
theories mostly in terms of configurations evolving in time (a
so-called `3+1' picture), he agrees that in the end the best way %%@
to
make sense of the `meshing' of different histories of %%@
3-dimensional
configurations is by thinking of each history as a foliation of a %%@
single
spacetime (cf.  Section 2.2.2.2 and 2.2.3).

 Second, Barbour is perfectly willing, indeed happy, to postulate %%@
less
structure for space than a fully-fledged manifold and metric. He %%@
has
hitherto postulated this rich structure for reasons, not of
philosophical conviction, but of pragmatism: so far, this %%@
structure
seems needed, if one is to get a precise Machian theory. But as %%@
he
reports (pp. 5, 349), he has work in progress (jointly with \'{O}
Murchadha) that aims to `gauge away length', i.e.  to weaken the
postulated metric structure so as to dispense with length, %%@
retaining
only the conformal structure (i.e. the structure of angles and
shapes).\footnote{These two qualifications are connected; part of %%@
the
motivation for conformal theories is that they promise to yield a %%@
preferred foliation of spacetime. For the joint work, cf. Barbour %%@
and \'{O}
Murchadha ([1999]).}

So to sum up this two-part answer:-- First, Barbour embraces the
development of field physics in that he seems content to %%@
postulate
space, and even to postulate that it has the rich structure of a
manifold with metric. But he is leary of spacetime, and would %%@
like to
postulate as weak a structure for space as possible, e.g. by
dispensing with lengths while retaining angles. Second, Barbour %%@
seeks
to understand space relationally; and in field theory this %%@
amounts to
thinking of spatial geometry as a matter of distances between
point-like parts of matter fields---a strategy that seems viable
outside general relativity.

\subsubsection{Machian theories}
In Barbour's technical work in classical physics, there have been %%@
two
main endeavours, the first leading in to the second.  First, he %%@
has
developed (together with Bertotti) novel Machian theories of %%@
classical
dynamics (both of point-particles and fields).  Second, he has %%@
given a
Machian analysis of the structure of general relativity. I will %%@
first
motivate, partly from a historical perspective, the simplest %%@
example
of the first endeavour, namely point-particles in a %%@
non-relativistic
spacetime.
That will introduce Barbour's brand of Machianism, and in %%@
particular
his central concept of instantaneous configurations. After %%@
presenting
this example in some detail, I will briefly describe how its main
ideas can be generalized to field theories and relativity, %%@
including
general relativity---and thereby touch on Barbour's second %%@
endeavour.\footnote{In this Section, I am much indebted to Oliver %%@
Pooley, and to his [2001], which gives many more technical %%@
details.}

\paragraph{2.2.2.1 Point-particles in a non-relativistic %%@
spacetime} Let us suppose that $N$ point-particles move in a %%@
spacetime that we assume is equipped
with an absolute simultaneity structure and Euclidean geometry in %%@
each
simultaneity slice; and maybe also with an absolute time metric. %%@
That
is already to assume a lot of abstract structure. But let us ask %%@
what
further information (facts) about the motion of the particles %%@
would be
acceptable ingredients in a Machian theory; in particular, what %%@
would
be acceptable initial data for the initial-value problem. Mach %%@
himself
gives no precise answer, though of course his relationism and
empiricism means he would favour facts about relational and/or
observable quantities. But as Barbour says (p. 71), in 1902
Poincar\'{e} did suggest a precise answer. He proposed that the %%@
theory
should take as initial data: (i) the instantaneous relative %%@
distances
between the particles (but {\em not} their positions in any %%@
putative
absolute space); and (ii) these distances' rate of change.
(Poincar\'{e}'s (ii) reflects his assuming an absolute time %%@
metric.)

One may well worry about the empiricist credentials of
Poincar\'{e}'s proposed initial data. For example: though %%@
instantaneous
inter-particle distances are no doubt relational, are they %%@
observable?
After all, ascertaining them takes time and a considerable amount %%@
of
theory: think of what it takes, in terms of laying out rods and
calculating. But it is best to postpone further discussion of %%@
such qualms to a
more general context (Section 2.2.3). Let us for the moment focus %%@
on
the technicalities of Machian theories.

The first point to note is that Poincar\'{e}'s proposal can be %%@
partly
motivated by considerations about geometric symmetries. Thus the
homogeneity and isotropy of Euclidean geometry suggest that the
instantaneous positions of the particles relative to any putative
absolute space, and the orientation of the whole system of {\em %%@
N}
particles relative to such a space, will be unobservable. More
precisely, they suggest that any such absolute positions and
orientation will have no effect on the subsequent evolution of
inter-particle distances. And this prompts the idea of %%@
formulating a
theory that describes the evolution of inter-particle distances %%@
(now
setting aside any empiricist qualms about these!)---while %%@
excluding
such positions and orientations from the initial data.\footnote{A
Machian should presumably resist the corresponding line of %%@
argument
appealing to kinematic symmetries. Familiarity with Galilean
relativity may suggest that a uniform motion of the whole system %%@
of
{\em N} particles relative to any putative absolute space will %%@
not
affect the subsequent evolution of inter-particle distances. But %%@
it
equally suggests that a rotation {\em might} do so. Thanks to %%@
Oliver Pooley for correcting me about this.}

Second, Poincar\'{e} saw that his proposed data, (i) and (ii), is
nearly, but not quite sufficient, to secure a unique future %%@
evolution
(`solve the initial-value problem') in {\em Newtonian} mechanics. %%@
We
can nearly, but not quite, express the usual Newtonian initial %%@
data
(positions and velocities in absolute space) in terms of this %%@
data. To
be precise: in order to rewrite the Newtonian initial data in %%@
terms of
inter-particle quantities, we need not only the inter-particle
distances and their first derivatives, but also three second
derivatives; or alternatively, one third derivative, and the %%@
values of
the total energy and the magnitude of the angular momentum.  This %%@
is a
very curious circumstance, especially since the need for just %%@
three
extra numbers is independent of the value of {\em N}. In any %%@
case, the
challenge to the Machian who accepts Poincar\'{e}'s proposal for %%@
what
is acceptable as initial data (and who believes in %%@
point-particles!)
is clear: find a theory for which this data {\em do} give a %%@
unique
future evolution---and show, if you can, its empirical %%@
equivalence, or
superiority, to Newtonian theory!

There is a history of such attempts; a tangled and ironic history
(unearthed over the last 20 years by Barbour himself, together %%@
with
other historians like John Norton). But I shall not linger on it, %%@
and
instead proceed directly to the eventual outcome, Barbour and
Bertotti's ([1982]) theory. As we shall see, this theory does %%@
away
with the absolute time metric; and its main ideas allow for
generalization to field theories and to relativistic theories.

The configuration space for the Newtonian mechanics of $N$
point-particles moving in an absolute three-dimensional space is
$3N$-dimensional Euclidean space $E^{3N}$.\footnote{Incidentally, %%@
the once-for-all labelling of the particles, and the
particle-labelling of the axes of any Cartesian coordinate system
$E^{3N}\rightarrow\mathR^{3N}$ need not involve assuming a %%@
so-called
`transcendental identity' for the particles: they might have %%@
different
masses.} Representing Newtonian time by the real numbers %%@
$\mathR$, a
possible history of the system is represented by a curve in %%@
$E^{3N}
\times \mathR$.  We can project any such curve  down into %%@
$E^{3N}$, the image-curve being the system's  orbit (in %%@
configuration space). It will be convenient to write the %%@
configuration space $E^{3N}$ as $Q$ (so that in later %%@
generalizations, we can again write the `absolutist' %%@
configuration space as $Q$); and to write its product with %%@
$\mathR$ as $QT$ (with $T$ for `time'). So we talk of projecting %%@
a curve in $QT$ down into $Q$.

We now introduce the idea of {\em relative} (instantaneous)
configurations of point-particles, and so the idea of a relative
configuration space (written `RCS')---the set of all such
configurations. (Barbour calls an RCS a `Platonia'; p. 44.)
Intuitively, a relative configuration is a specification of all %%@
the
inter-particle distances (and so of all the angles) at some %%@
instant,
without regard to (a) where the system as a whole is in absolute
space, nor to (b) how it is oriented, nor to (c) its handedness. %%@
We
can formally define relative configurations in terms of an %%@
equivalence
relation on the `absolutist' configuration space $Q = E^{3N}$. %%@
Let us
say that two points $x, y
\in E^{3N}$ are equivalent, $x \sim y$, if for each pair of
point-particles, $m_i, m_j$ say, $x$ and $y$ represent the same
inter-particle distance between $m_i$ and $m_j$. In other words, %%@
if we
think of $x$ and $y$ as specifying polyhedra $\pi(x)$ and %%@
$\pi(y)$ in
physical space (with each vertex labelled by its particle), we %%@
define:
$x \sim y$ iff $\pi(x)$ is transformed to $\pi(y)$, with the
$m_i$-vertex in $\pi(x)$ being transformed to the $m_i$-vertex in
$\pi(y)$, by some element of the improper Euclidean group on %%@
$E^3$.
(`Improper' so as to identify oppositely handed polyhedra, i.e.
incongruent counterparts.)  The quotient space $E^{3N}/\sim$
consisting of the equivalence classes is the relative %%@
configuration
space. We can project any curve in $E^{3N}$ down into this RCS. %%@
It
will be convenient to write this RCS as $Q_0$, so as to have a
convenient notation later for other RCSs; so we speak of %%@
projecting a
curve from $Q$ to $Q_0$.

As discussed above, the initial data for the Newtonian mechanics %%@
of
point-particles consists of a bit more than the inter-particle %%@
distances and
their rates of change.  In terms of the configuration spaces we %%@
have
just introduced, this means that an initial point in $Q_0$, %%@
together with
an initial vector at that point (representing inter-particle %%@
distances' rates of change), are not in general sufficient to
pick out the unique future evolution prescribed by Newtonian
mechanics; (where the curve representing this evolution is %%@
projected
from $QT = E^{3N}
\times \mathR$ first to $Q = E^{3N}$ and then to $Q_0$). But if %%@
we specify a small amount of further information, then we can
succeed; and such information can be specified in various ways. %%@
Of
course from a Newtonian perspective, there will be little to
choose between these ways. Any such specification will seem an
artificial form of initial data, the need for which arises from %%@
our
having chosen to follow Poincar\'{e} in focussing on purely %%@
relational
data.

But from a Machian perspective, the various ways to secure from
relational initial data a unique trajectory (which in principle %%@
need
not be a possible Newtonian one) are of central interest. And it %%@
turns
out that we can secure a unique trajectory with a specification %%@
that
includes, not an initial vector at the initial point in $Q_0$, %%@
but
merely an initial direction: that is, it includes the %%@
inter-particle
distances' rates of change only upto an overall scale factor, %%@
which would set
`how fast the absolute time elapses'.  A bit more precisely: it %%@
turns
out that if we impose zero total angular momentum and a fixed %%@
energy,
then an initial point in $Q_0$, and an initial direction at that
point, secures a unique trajectory which is a solution of the %%@
Newtonian theory.

So the Machian aims to find a theory that has just these %%@
solutions as
all its solutions. Any such theory will deserve to be called
`relational' in the sense that its dynamics can be formulated %%@
wholly
in terms of the RCS $Q_0$. And it will deserve to be called %%@
`timeless',
in that there is no time metric in $Q_0$; rather, as we will see %%@
in a
moment, the time metric is definable from the dynamics. Barbour %%@
and
Bertotti ([1982]) state such a theory. I first state the three
fundamental ideas of the theory, as (i) to (iii); and then state %%@
results
arising, as (1)-(3).

(i) We begin by postulating an RCS $Q_0$ of relative %%@
configurations of
$N$ point-particles (distinguished once-for-all from each other, %%@
e.g.
by having distinct masses) and we require each configuration to %%@
obey
3-dimensional Euclidean geometry. This requirement means that if %%@
we
embed a relative configuration in Euclidean space $E^3$, the
particles' $3N$ coordinates in a Cartesian coordinate system will %%@
not
be independent. More generally, it gives the RCS a very rich
structure; (discussed on e.g. pp. 40-46, 71-86). 

(ii) Suppose given an (isometric) embedding $\theta$ of two %%@
relative
configurations $x, y$ into $E^3$. Given an ordering of the %%@
particles,
this induces an embedding of $x, y$ as points in $E^{3N}$,
$\Theta:x,y\mapsto\Theta(x), \Theta(y) \in E^{3N}$. Then the %%@
usual
Euclidean metric $d^E$ on $E^{3N}$ defines a distance depending %%@
on
$\Theta$, and so $\theta$, between $x$ and $y$: $d_{\theta}(x,y) %%@
:=
d^E(\Theta(x),\Theta(y))$. Given a Cartesian coordinate system on
$E^{3N}$, this distance can be expressed in the usual way as the
square root of a sum of squares of differences of coordinate %%@
values.
Of course, $d_{\theta}(x,y)$ gives no measure of the intuitive
similarity of $x$ and $y$ since $\theta(x)$ and $\theta(y)$ can %%@
be
arbitrarily far apart in $E^3$, and so $\Theta(x)$ and %%@
$\Theta(y)$ can
be arbitrarily far apart in $E^{3N}$, no matter how similar $x$ %%@
and $y$ are.

Barbour and Bertotti define a measure of intuitive similarity, %%@
i.e. a
metric on $Q_0$, by considering for any $x$ and $y$, all possible
embeddings $\theta$ of $x$ and $y$ into $E^3$, and taking the %%@
minimum
value of a metric, ${\bar {d_{\theta}}}(x,y))$ say, which is just %%@
like $d_{\theta}$ except that each squared difference of %%@
coordinate values is weighted by the mass of the particle %%@
concerned; (cf. pp. 115-118). Think of putting an
overhead transparency with $N$ differently coloured dots, on top %%@
of
another transparency with $N$ such dots, in such a way as to %%@
minimize
the weighted sum of the squared distances for red-to-red, %%@
green-to-green etc., with the weights given by a mass associated %%@
with each dot.
Some terminology: the minimization procedure is called `best
matching'; the resulting metric is called the `intrinsic %%@
difference'
(or infinitesimally, `differential') between the relative
configurations $x$ and $y$; and the theory we are constructing is
called `intrinsic dynamics'.

This metric also provides a solution to a problem that arises %%@
once we
think of each relative configuration as in its own instantaneous %%@
space
(a copy of $E^3$): the problem of identifying spatial points (not
point-particles) between two such spaces---a problem which %%@
Barbour
calls the `problem of equilocality'. The solution is that two %%@
spatial
points in the instantaneous spaces of two relative configurations %%@
$x,
y$ are equilocal if they have the same coordinate values in those
Cartesian coordinate systems on $E^3$ that minimize ${\bar %%@
{d_{\theta}}}(x,y))$.
More terminology: two relative configurations with their spatial
points thus identified are called `horizontally stacked'. The %%@
idea is
that, as usual in spacetime diagrams, the vertical dimension (`up %%@
the
page') represents time, so that `horizontal stacking' refers to
placing relative configurations relative to one another in the
horizontal dimensions while stacking them in a vertical pile %%@
(i.e. in
the time dimension).\footnote{The idea that one should identify %%@
places
across time in such a way as to minimize the resulting motion of
bodies has also been discussed in the philosophical literature: %%@
for
example, {\em pro} Peacocke ([1979], p.  50-51) and {\em con} %%@
Forbes
([1987], pp.300-304).}

(iii) Barbour and Bertotti now combine the intrinsic metric with %%@
some
remarkable work on Newtonian point-particle mechanics by Jacobi.
Jacobi showed that for a conservative Newtonian system the orbit %%@
in
the absolute Newtonian configuration space, i.e. $Q = E^{3N}$, %%@
can be
found from a variational principle defined solely on $E^{3N}$ %%@
(instead
of $QT = E^{3N} \times \mathR$ as in Lagrange's approach). %%@
According to this
principle (now called `Jacobi's principle'), the `action' to be
extremized is the integral along a curve in $Q$ of the square %%@
root of
$T$ times $E-V$, where $T$ is a generalized kinetic energy, $E$ %%@
is the
total energy and $V$ is potential energy. That is, we are to %%@
extremize
$I_{\rm Jac} := \int\/d\lambda \surd[T.(E-V)]$; where $\lambda$ %%@
gives
an arbitrary parameterization along the curve, so that there is %%@
no use
of a time metric. (Thus each value of the total energy $E$ %%@
specifies a
different variational principle.) 

The idea of intrinsic dynamics is to replace the generalized %%@
kinetic
energy $T$ in $I_{\rm Jac}$ by an analogue, $T_{\rm Mac}$, that %%@
uses
the intrinsic differential; ($T_{\rm Mac}$ becomes $T$ in %%@
horizontally
stacked coordinates). More precisely, the theory postulates that %%@
the
system's evolution is given by a variational principle on $Q_0$, %%@
viz.
by extremizing the expression obtained by substituting $T_{\rm %%@
Mac}$
for $T$ in $I_{\rm Jac}$. Let us call this expression $I_{\rm %%@
Mac} :=
\int\/d\lambda \surd[T_{\rm Mac}.(E-V)]$. (So, since the %%@
intrinsic
differential itself involves an extremization, there is a double %%@
variation to
extremize $I_{\rm Mac}$.)

Thus defined, intrinsic dynamics yields the following results. %%@
(1) All
curves in $Q = E^{3N}$ that project down to the same curve in %%@
$Q_0$
have the same value for $I_{\rm Mac}$. So intrinsic dynamics'
postulate that evolution is given by extremizing $I_{\rm Mac}$ %%@
really
is, as desired, a variational principle on $Q_0$. We have a %%@
dynamics
that specifies unique curves from data in the RCS.

(2) If we use spatial coordinates corresponding to horizontal %%@
stacking,
then extremizing $I_{\rm Mac}$ reduces to extremizing $I_{\rm %%@
Jac}$.

(3) We can recover the familiar Newtonian time metric, as %%@
follows. The
equations of motion given by intrinsic dynamics become the %%@
familiar
Newtonian ones if we choose the arbitrary parameter $\lambda$¸ %%@
along
curves in $Q_0$ in such a way that $T = E - V$. And analogously %%@
to the
terminology of horizontal stacking, we say that this choice of
$\lambda$ (this assignment of time differences between %%@
configurations
in a possible history) `vertically stacks' the configurations.  %%@
So
`vertical stacking' refers to how far apart the configurations %%@
should
be placed in the time dimension.

This choice of $\lambda$, which recovers the Newtonian time, is %%@
made
once-for-all for the entire system (the universe). But intrinsic
dynamics also provides an explanation of why subsystems that are
effectively isolated from one another behave as if they evolved
according to a common time parameter. For it turns out that if %%@
for
such subsystems, we make the corresponding choice of $\lambda$ %%@
(i.e.
we again impose $T = E - V$), then their different times `march %%@
in
step' with each other. This is a significant
result just because the observed fact, that mutually isolated
subsystems of the universe behave as if they evolved according to %%@
a
common time parameter, is so striking.\footnote{Or rather, it is
striking once it is pointed out! I suppose Newton may well have %%@
had it
in mind when he wrote the famous remark in the Scholium that %%@
absolute
time `from its own nature flows equably without relation to %%@
anything
external.' But nowadays, after the rise of relativity, this %%@
remark is
most commonly read, not as a statement of the absoluteness
(system-independence) of the temporal metric (durations), but as %%@
a
statement of the absoluteness (frame-independence) of
simultaneity---which presumably Newton did {\em not} have in %%@
mind!}

Two final comments on this theory, which fill out its claim to be %%@
a
`timeless' as well as `relational' theory. First, these results,
especially (2) and (3), give a clear sense in which the absolute %%@
time of
orthodox Newtonian mechanics emerges within intrinsic dynamics. %%@
And as
I said in Section 2.2.1, this is not emergence in the typical
philosophical sense of `approximate validity', or some other weak
analogue of reduction. The Newtonian time parameter is exactly
definable within intrinsic dynamics. More precisely, in a
formalization of intrinsic dynamics using a sufficiently powerful
background logic (including e.g. multivariate calculus), the %%@
Newtonian time
parameter would be finitely definable (not just
supervenient).\footnote{Agreed, we can also write the usual %%@
Newtonian
theory in a generally covariant form in which we can explicitly %%@
define
absolute time in terms of the covariant timelike vector field %%@
$t_a$;
but in this case, the absolute time is clearly `already there' %%@
and
`non-dynamical' in the given formalism---and so not emergent.}

Second, the idea in result (3), of fixing the time metric by %%@
means of
the dynamical evolution of the whole system, is reminiscent of a
development in conventional, i.e. Newtonian, astronomy; %%@
(discussed by
Barbour, pp. 97-99, 104-108). In Newtonian astronomy, the %%@
assumption
that there is an absolute time parameter governing all bodies'
evolutions, gives no guarantee that any single body's motion is
exactly periodic with respect to it. In particular, there
is no guarantee that solar time and sidereal time (given by the %%@
return
to a `position' of the sun, and of a given star, respectively) %%@
measure
absolute time (i.e. stay in step with it).  But remarkably---and %%@
very
luckily for the development of physics!---solar and especially
sidereal time proved sufficiently accurate until about 1900, when %%@
the
errors due to inter-planetary interactions began to show up. In %%@
order
to obtain still greater accuracy, astronomers then resorted to %%@
the idea of
assuming that Newtonian mechanics with some absolute time %%@
parameter
(called `ephemeris time') governed the solar system {\em as a %%@
whole};
they then deduced where the various heavenly bodies would be at %%@
given
values of the assumed parameter---in fact using the Moon as the %%@
`hand'
on the face of the `clock'. This idea of ephemeris time is %%@
obviously
similar to the idea in result (3).\footnote{The history of this %%@
topic goes back to Ptolemy! Cf. Barbour ([1989]), pp. 175-183.}

\paragraph{2.2.2.2 Field theory and relativity} I turn to %%@
sketching how these ideas of intrinsic dynamics are adapted to %%@
classical field theory and relativity, including general %%@
relativity. But before giving details, I should first recall %%@
(from Section 2.2.1)
that Barbour wants to understand space relationally; and so he %%@
hopes
to understand field theories as postulating point-like parts of %%@
matter
fields and treating spatial geometry in terms of distances %%@
between
these point-like parts. Though this stance will not much affect %%@
the
formal details to follow,\footnote{But this is partly a matter of
suppressing subtleties for the sake of expository clarity and %%@
brevity.
In particular, I have suppressed discussion of the relationist's %%@
need
to deal with symmetric configurations, where the field takes %%@
identical
values at intuitively distinct locations.} it is interpretatively
significant---not least because it obviously looks less plausible %%@
for
theories like general relativity that have a dynamical metric,
interacting `on all fours' with matter fields.

But let us start with the easier case of a non-dynamical metric. %%@
For a
field theory using a non-dynamical Euclidean spatial structure %%@
(with
absolute time, or in special relativity), the RCS, which I again %%@
call
$Q_0$, will consist of all possible instantaneous relative field
configurations in Euclidean space. As in the point-particle case,
$Q_0$ can be obtained as a quotient of a configuration space $Q$
consisting of `absolute' field configurations, quotienting by the
action of the Euclidean group. But this $Q_0$ is a much more
complicated RCS than that in the point-particle case. In %%@
particular,
since there are infinitely many independent ways that two field
configurations can differ, this RCS will be infinite-dimensional. %%@
So
let us only aim to consider the simplest possible case: a scalar %%@
field
with compact spatial support. In this case, one can define `best
matching' and so an intrinsic metric on the configurations.
Furthermore, in the relativistic case, one can pass from a
Lorentz-invariant action principle of the usual Lagrangian kind %%@
(i.e.
defined on a relativistic $QT$, extremizing a 4-dimensional %%@
integral
$\int d^4x \dots$) to a Jacobi-type principle on $Q$; and thereby %%@
to a
Machian action principle, defined on $Q_0$ and extremizing %%@
$I_{\rm
Mac}$.

Just as in the case of point-particle mechanics: all curves in %%@
the absolute
configuration space $Q$ that project to the same curve in $Q_0$ %%@
have
the same value for $I_{\rm Mac}$. So the Machian action principle
really is a variational principle on $Q_0$. In this sense, we %%@
have a
genuinely relational field theory. 

Furthermore, the original action principle being %%@
Lorentz-invariant
secures that the corresponding intrinsic dynamics is
special-relativistic in the sense that: any of the theory's
dynamically possible curves in $Q_0$, once it is horizontally and
vertically stacked, represents a (4-dimensional) solution of the
original (orthodox) special-relativistic field theory.

But there is also another sense of being special-relativistic %%@
that we
need to consider; (which is closer to the idea of a principle of
relativity). Namely: suppose we apply a passive %%@
Lorentz-transformation
to such a 4-dimensional solution, and analyse it as a sequence of
instantaneous 3-dimensional configurations in the new frame. This
sequence will of course define a different curve in $Q_0$ from %%@
that defined by
the description in the original frame. So let us ask: is this new
curve in $Q_0$ also dynamically allowed by the corresponding %%@
intrinsic
dynamics? That is, is it a geodesic of the variational principle %%@
on
$Q_0$? The answer is Yes provided that the principle satisfies %%@
some
restrictions. That is again a satisfying result; but it
prompts an interpretative question about the sense in which the
`3+1'-perspective favoured by Barbour can be taken as %%@
fundamental.
However, I will postpone this question until after discussing %%@
general
relativity from an intrinsic dynamics perspective (i.e. until the %%@
next
subsection), since the very same question also arises for general
relativity.

Despite having a dynamical metric, general relativity can be
formulated in a `3+1' form, i.e. in terms of the coupled %%@
evolution
over time of spatial geometry and matter-fields. Though Einstein
himself always thought of it instead in spacetime terms %%@
(epitomized by
his field equations applying to each spacetime point), the `3+1'
approach, called `geometrodynamics', was developed in the 1950s %%@
and
1960s by Dirac, Wheeler and others. In this approach, the theory %%@
is expressed not by Einstein's field equations, but by the %%@
momentum and Hamiltonian constraint equations mentioned in %%@
Section 1. But note that in some ways, this approach represents %%@
only part of general relativity. For it involves fixing the %%@
global topology of spacetime to be a product of the topologies of %%@
some spatial 3-manifold and of the real numbers (representing %%@
time); and it therefore forbids the kind of topology change that %%@
general relativity allows---and that occurs or at least seems to %%@
occur, for instance in black holes.

Nevertheless, suppose we adopt this approach and again consider %%@
the simplest possible case. We begin by representing space as a %%@
compact 3-manifold
$\cal M$, so that spatial geometry is given by a Riemannian %%@
metric $h$
on $\cal M$; and let us ignore matter fields, so that we are %%@
concerned
with the time-evolution of geometry alone---so-called `pure (i.e.
vacuum) geometrodynamics'. From the traditional relationist %%@
perspective, this last
will surely be an unacceptable simplification. But let us here %%@
set
aside this qualm, since: (i) in fact Barbour himself willingly
discusses pure geometrodynamics (both in this book and his %%@
technical
papers), though it is hard to square with the idea (cf. Section %%@
2.2.1) of treating spatial points as point-like parts of a %%@
matter-field; and (ii) relationism aside, the conceptual issues %%@
about
geometrodynamics arise in an undistorted but technically simpler %%@
form,
for the compact manifold and vacuum case. Then the obvious
configuration space to consider is the set $Riem({\cal M})$ of %%@
all
Riemannian metrics on $\cal M$. Again, this is a very complicated
space; in particular it is infinite-dimensional.

 But in fact, much technical work in geometrodynamics uses %%@
another
space, the quotient space of $Riem({\cal M})$ under the action
of spatial diffeomorphisms. That is, one defines an equivalence
relation on $Riem({\cal M})$, saying that $h \sim h'$ iff $h, h'$ %%@
can
be mapped into one another by a diffeomorphism of $\cal M$. The
resulting quotient space, which we can write as $Riem({\cal
M})/Diff({\cal M})$, is called `superspace'.  The use of this %%@
quotient
space corresponds to imposing the momentum constraint; and it is %%@
of course in the spirit of the hole argument (mentioned in
Section 2.1.1), since the conclusion of that argument is %%@
precisely that
we should attribute no physical significance to {\em which} point %%@
of
$\cal M$ has a given array of metrical properties. So general
relativity, once formulated as geometrodynamics on superspace, %%@
already
seems close to the spirit of Barbour's views, with $Riem({\cal %%@
M})$
serving as the `absolutist' $Q$, and $Riem({\cal M})/Diff({\cal %%@
M})$
as Barbour's preferred RCS $Q_0$.

Indeed, Barbour shows that there is a stronger connection; (this %%@
is
his Machian analysis of general relativity, pp. 170-177).  Recall %%@
how
intrinsic dynamics for point-particles used the metric on the
absolutist configuration space $Q = E^{3N}$, together with the %%@
idea of
best matching, to define a metric, and thereby a variational %%@
principle
for dynamics, on the RCS $Q_0$. In a similar way, an intrinsic
dynamics approach to geometrodynamics will aim to have a metric %%@
on
$Riem({\cal M})$ induce a metric on $Riem({\cal M})/Diff({\cal %%@
M})$,
the geodesics of which will represent the dynamically possible
histories of 3-geometries. Barbour shows\footnote{Of course not %%@
here!
Cf. e.g. his ([1994]), pp. 2864-2869.} that this approach leads %%@
to a
certain general form of the `action' to be extremized on
superspace---and it turns out that under certain conditions, the
orthodox 4-dimensional action of general relativity can be put in %%@
a
form---the Baierlein-Sharp-Wheeler form---that instantiates %%@
Barbour's
general form. (The passage from the orthodox 4-dimensional action
principle to the Baierlein-Sharp-Wheeler form is analogous to the
corresponding passage in special relativity: from $QT$ to $Q$ to
$Q_0$.) Furthermore, in order to obtain the %%@
Baierlein-Sharp-Wheeler
form, one does not need to impose restrictions on general %%@
relativity,
analogous to the fixed energy and zero angular momentum %%@
restrictions
that earlier had to be imposed, in order to get the same solution
space as given by intrinsic dynamics. Thus Barbour says that %%@
general
relativity is already (i.e.  without extraneous restrictions %%@
being
needed) a `frameless' theory.

Besides, general relativity is a timeless theory, in the same %%@
sense in
which we saw non-relativistic point-particle intrinsic dynamics %%@
is:
viz.  that the time metric is explicitly definable from the %%@
dynamics.
For starting with Barbour's general form for the Machian
geometrodynamic `action', we can choose the parameter $\lambda$ %%@
on
geodesic curves by imposing the local analogue of the global %%@
condition
$T = E - V$ that we chose in the point-particle case. And if in
particular, we do this starting with the Baierlein-Sharp-Wheeler
action of general relativity, we recover precisely general
relativity's usual notion of proper time. So as Barbour puts it, %%@
we
can think of proper time as a local ephemeris time.

Finally, I turn to the topic of different foliations of a single
spacetime. I already discussed this briefly for special %%@
relativity,
but for general relativity the situation is more interesting, and
indeed more positive. Suppose we again start with Barbour's %%@
general
form for the Machian geometrodynamic `action', and choose %%@
$\lambda$ in
the way just discussed to construct a 4-dimensional spacetime %%@
from an
extremal sequence of 3-geometries (i.e. a sequence that %%@
extremizes
Barbour's form).  Now consider a different foliation of the %%@
spacetime
just constructed, but with the same end-points (initial and final
3-surfaces); and consider the sequence of 3-geometries thus %%@
defined.
Now let us ask: is this different sequence also a solution to our
Machian theory?  Does it also extremize Barbour's form? The %%@
answer in
general is `No'. But general relativity is one of a very small %%@
class
of theories for which the answer is `Yes'. Indeed, given some
simplicity conditions, general relativity is the only such %%@
theory.

\subsubsection{Assessing intrinsic dynamics}
I shall make three general comments about Barbour's theories of
intrinsic dynamics; proceeding from `the good news' through to
`the bad news'. I shall first state what I take to be these %%@
theories'
main novelty and merit. Then I shall discuss the empiricist %%@
credentials
of Barbour's instantaneous configurations. Finally I shall raise %%@
a
difficulty about relativity. It is a difficulty that Barbour
recognizes: and it will lead to the discussion of quantum theory %%@
in
Section 3.

(1) The main novelty and merit of these theories is of course %%@
their
use of relative configurations, and their timelessness, in the %%@
sense
that the time metric is wholly defined by the dynamics. It is %%@
indeed
remarkable that there are classical theories of mechanics and %%@
field
theory with these features, that are nevertheless close cousins,
theoretically and observationally, of the usual ones. Since I %%@
will
discuss configurations in (2) (and since time is our main %%@
topic!), I
shall here just make one remark about timelessness, supplementing %%@
the
end of Paragraph 2.2.2.1. The familiarity of classical theories %%@
`with
time'---meaning, not absolute simultaneity, but that time is %%@
treated
independently of the matter content of universe---should not %%@
blind us
to the timeless alternative; (more precisely: blind us, once we %%@
set
aside general relativity, whose dynamical metrical structure is
nowadays also familiar). Indeed, as Barbour himself stresses, %%@
these
theories' success depended on a remarkable and fortuitous
circumstance: that until 1900, the frame defined (roughly %%@
speaking)
spatially by the fixed stars and temporally by the rotation of %%@
the
Earth, was within observational limits an inertial frame in which
Newton's laws held good. As Barbour puts it (p. 93), there really %%@
seemed to be
a `clock in the sky'.

(2) We have seen that the main ingredient of the various theories %%@
of
intrinsic dynamics are 3-dimensional instantaneous %%@
configurations. So
it is reasonable to ask how well the notion of configuration
accords with the tradition of Mach, or of relationism in general.
Though the exact meaning of `configuration' of course differs %%@
from one
theory to another, we can reach some general conclusions.

The Machian or relationist must favour configurations that %%@
involve
only `observable' and/or `relational' facts or physical %%@
quantities.
Maybe we can rest content that in our first case, that of
point-particles, the inter-particle distances are both observable %%@
and
relational. But in general, the two notions may well come apart.
Instantaneous facts, and similarly physical quantities whose %%@
values
encode a configuration, can surely be observable but %%@
non-relational;
(e.g. the absolute temperature of a state of a matter field at a
spacetime point or local region). And vice versa: such facts and
quantities can be relational but unobservable; (e.g. the distance
between simultaneous spacetime points). In view of this, one %%@
might
suggest that for a truly Machian or relational theory, the facts %%@
and
quantities encoding a configuration should be both relational and
observable.

Here there are two points to be made: the first, (a), is a %%@
general
problem; and the second, (b), concerns Barbour's own views about %%@
what
counts as a Machian configuration.

(a) There remains the problem I mentioned at the start of %%@
Paragraph
2.2.2.1, about the empiricist credentials of the very idea of a
configuration: ascertaining even `elementary' facts such as
inter-particle distances takes time and a considerable amount of
theory. Besides, one can perfectly well press Section 2.1.2's
consistency problem for configurations, just as for other aspects %%@
of
physical geometry. Accordingly, an empiricist should think twice %%@
about
taking instantaneous configurations as the basic ingredients of a
dynamics.

 A similar point applies to our everyday impression that we see
the present state of objects situated across a stretch of space.
For Barbour, this impression is important for motivating
configurations as the basic ingredient of dynamics. For example,
he writes:
\begin{quote} Is not our most primitive experience always that we
seem to find ourselves, in any instant, surrounded by objects in
definite spatial positions? Each experienced instant is thus of
the nature of an observation .... Moreover what we observe is
always a collection, or totality, of things. We see many things
at once. (p. 265)
\end{quote}
\noindent This impression is indeed striking. But I
doubt that it has any heuristic value for dynamics, or indeed for
physics in general. For it clearly arises from the fact that
perception is usually sufficiently rapid, compared with the
time-scale on which macroscopic objects change their observable
properties, that we can take perception to yield information
about objects' present states, i.e. their states at the time of
our perceptual judgment.  This fact is worth remarking; it even
has some philosophical repercussions (Butterfield [1984]). But it
is also clearly a contingent fact about our perceptual system,
albeit an adaptive, and no doubt evolved, one---and unlikely to
be a clue to the structure of dynamics.

(b) To judge from the theories that Barbour himself has %%@
constructed or
studied, and called `Machian', he sets much less store by
observability than by being relational.  Though he wants to %%@
understand
space relationally (as discussed in Section 2.2.1), he %%@
nevertheless
counts as Machian a theory whose initial data include the
instantaneous state of a scalar field; which is hardly %%@
observable,
being defined at every point on a spacelike surface. Furthermore %%@
he
says the same for pure geometrodynamics. That is, he counts as %%@
Machian
a theory whose initial data specify a Riemannian geometry, but %%@
zero
matter fields, on a 3-dimensional manifold.

On the other hand, it is of course important to Barbour that the
configurations, and the quantities that encode them, be `as %%@
relational
as possible'.  This is clear from the theories discussed above. %%@
And as
I mentioned in Section 2.2.1, he is currently working on %%@
`conformal'
Machian theories in which there is no length scale, so that for %%@
the
case of point-particles, three particles can form a unique %%@
equilateral
triangle---and in general, not only are congruent triangles of
Euclidean geometry, differing from one another in their location %%@
and
orientation in Euclidean space, to be identified, but so also are %%@
the
{\em similar} triangles.

(3) In Paragraph 2.2.2.2, I reported that intrinsic dynamics can
respect relativity's freedom to foliate a spacetime in countless
different ways, in the following sense. For both special and %%@
general
relativity, if we are given a spacetime constructed from one %%@
solution
of intrinsic dynamics (one extremal sequence of 3-dimensional
configurations) and we consider any foliation of the spacetime, %%@
then
the resulting sequence of configurations on that foliation is %%@
also a
solution of intrinsic dynamics. I also mentioned that though %%@
these are
satisfying results, they prompt the question whether the
`3+1'-perspective favoured by Barbour can be taken as %%@
fundamental.\footnote{I set aside the more basic physical %%@
objection mentioned in Paragraph 2.2.2.2, that the `3+1' %%@
perspective captures only part of general relativity, since it %%@
forbids topology change.}
This is the question to which I now turn. 

Of course, `fundamental' can mean different things, and the
verdict about this question is likely to depend on the meaning %%@
chosen.
Here it will be enough to indicate two reasonable connotations of
the word, and to remark that one of them implies that a `Yes' %%@
answer
(i.e. saying that configurations are fundamental) would conflict
with another desideratum, determinism---the conflict arising in a %%@
way 
analogous to the hole argument.

According to intrinsic dynamics, instantaneous configurations are
certainly fundamental in the senses that comparison of their %%@
internal
structures defines equilocality and the temporal metric, and that %%@
the
dynamical law is a geodesic principle on the RCS, so that a %%@
geodesic
completely describes a possible history of the universe. But we %%@
need
to be more precise about `complete description'. Do we mean that %%@
such
a geodesic determines a history though not {\em vice versa}, thus
allowing that distinct geodesics might determine the same %%@
history? If
so, then the different foliations of a relativistic spacetime %%@
will
correspond to different geodesics describing the same possible
history. Or do we mean, more strongly, that such a geodesic %%@
describes
a history of the universe, not only completely but also
non-redundantly (in physics jargon: with no gauge-freedom)---so %%@
that a
history corresponds to a unique geodesic? If so, then the %%@
countless
different foliations of a relativistic spacetime will commit us %%@
to
saying that such a spacetime contains countless different %%@
histories.
This conclusion seems implausible.  Furthermore, since in general
relativity (or even special relativity if we allow a foliation's
slices not to be hyperplanes), two foliations can match upto a %%@
slice
and diverge thereafter, this conclusion implies indeterminism of
histories, in a manner reminiscent of the hole %%@
argument.\footnote{This
`hole argument for Barbour' is due to Pooley ([2001]), Section %%@
3.5. Pooley concedes that Barbour could reply to this argument %%@
that his recent work on conformal theories promises to give a %%@
preferred foliation; cf. footnote 11.}

Barbour is aware of this tension between the spacetime viewpoint %%@
of
relativity and his advocacy of configurations and a `3+1' %%@
viewpoint;
though he does not express it in these philosophical terms (in %%@
this
book or elsewhere). His answer, at least in this book, is to %%@
concede
that for classical physics, the tension is resolved in favour of %%@
the
spacetime viewpoint. He writes:
\begin{quote} If the world were purely classical, I
think we would have to say ...that the unity [of spacetime, %%@
which] Minkowski
proclaimed so confidently is the deepest truth of spacetime. %%@
(p.180)
\end{quote}
\noindent But as the antecedent suggests (and the rest of the %%@
passage, pp.
180-181, makes clear), he also believes that in a quantum world, %%@
this
tension is resolved in the other way: his advocacy of %%@
configurations
wins---cf. Section 3.

\section{The End of Time?}
I turn to Barbour's view that there is no time. In this Section,
I will emphasise philosophy not physics; and not just for lack
of space, and as befits this journal---there are two other %%@
reasons. As I said in Section 1, Barbour's denial of
time can be to a large extent understood and assessed
independently of physics, in particular the technicalities of
quantum geometrodynamics, i.e. the approach to
quantum gravity which Barbour favours.\footnote{In particular, %%@
the denial of time is not just the claim that the temporal metric %%@
is emergent {\em a la} Section 2.2.2: which would anyway no more %%@
be a denial of time than the reduction of light to %%@
electromagnetic waves is a denial of light.} Also, he does not %%@
relate his denial of
time very precisely to various philosophical positions; in
particular, the tenseless (`B-theory') of time. So to understand
his denial, we need to be careful.\footnote{Besides, the issues
are in any case confusing! As we shall see, even such an exact
thinker as John Bell was similarly imprecise, when he formulated
a `denial of time', in an essay which strongly influenced
Barbour.} 

It will be clearest to begin in Section 3.1 by stating three %%@
possible
meanings of the slogan that `time is unreal', which all make %%@
sense
more or less independently of physics.  More precisely: to the %%@
extent
that physics is relevant, one can ignore the peculiarities of %%@
quantum
theory, and focus entirely on classical physics. Then I will be %%@
able
to state the main idea of Barbour's own denial of time. It will %%@
give
a fourth meaning of the slogan. It has some similarities to the %%@
first
and third meanings; but it also has distinctive features of its %%@
own,
related to Barbour's notion of a special type of configuration %%@
called
a `time capsule'. Then in Section 3.2, I will turn to quantum %%@
physics.
This fills out the discussion of Barbour's denial of time in two %%@
main
ways. First, Barbour urges, on the basis of some previous %%@
orthodox
work, that in quantum physics time capsules get high probability
(according to the quantum state). Second, he discusses the %%@
problem of time of quantum geometrodynamics. He argues that the %%@
best solution to this
problem is to deny time in the way he proposes.

\subsection{Time unreal? The classical case}
I will first state three possible meanings of the slogan that %%@
`time is
unreal'.\footnote{Of course there are yet other meanings; for
example, McTaggart's suggestion of a `C-series'.} To
philosophers, the first two (Section 3.1.1) are familiar broadly %%@
defined
positions: the denial of temporal becoming, which I will call
`detenserism' (an ugly word, but only one); and the claim that %%@
only the present is real, which is nowadays called `presentism' %%@
(also ugly, but so be it!). The
third, which I call `Spontaneity', is probably unfamiliar %%@
(Section 3.1.2). But once
we have grasped it, we can state the essential idea of Barbour's
own denial of time, for the context of classical physics (to %%@
which
this Subsection is confined). In effect the idea is a hybrid of
Spontaneity and a Lewis-like modal realism (Section 3.1.3). 

It will be clear from this Subsection's discussion that {\em all} %%@
of
the classical theories mentioned in Section 2, whether
traditional or Machian, fit very well with the first meaning of
time being unreal, i.e.  detenserism. That is a familiar point, %%@
and I will not belabour
it.\footnote{This familiarity is one reason why it is tempting to
misread Barbour's denial of time as merely denying temporal %%@
becoming.} But more importantly, it will also be clear that {\em %%@
none} of
Section 2's classical theories give any support to the other
meanings of time being unreal. And here by `other meanings', I
intend not only the second and third, presentism and Spontaneity;
but also the fourth---Barbour's own denial of time. That denial
is most easily grasped in the context of classical physics (as
are the other meanings). But as we shall see in Section 3.2, %%@
Barbour's
reasons for believing it, rather than just formulating it, come
entirely from quantum physics. Indeed, they come from the problem
of time in quantum gravity, in the form that problem 
takes within quantum geometrodynamics.

\subsubsection{Detenserism and presentism} Saying that time is %%@
unreal could mean denying so-called
`temporal becoming', i.e. advocating the `block universe' or %%@
`B-theory
of time'. The idea is that past and future things, events and %%@
states
of affairs (or however one conceives the material contents of
spacetime) are just as `real' as present ones. Abraham Lincoln is %%@
just
as real as Bill Clinton, just as Venus is just as real as Earth:
Lincoln is merely temporally `far away from us', just as Venus is
spatially far away.  Similarly for my first grandchild, supposing %%@
I
have one: where this caveat simply reflects the fact that it is %%@
hard
to know about the future (even harder, perhaps, than it is to %%@
know
about the past)---not that the future, or its material contents, %%@
is of
some different ontological status than the present or past.

On the other hand, saying that time is unreal could mean
`the opposite', i.e. presentism. The idea is that only the %%@
present
is real: past and future things etc. are unreal. Here, I say
`things etc.' for simplicity: as regards the main idea of
presentism, it does not matter how you conceive the material
contents of spacetime---though of course in more precise %%@
versions, it can
matter.

Thus the debates about these two positions turn on the contrast
between the real and the unreal: no wonder both are %%@
controversial! I
shall not try to contribute to them, but I need to stress three
points: about dangers of ambiguity, about modality and about
semantics, respectively. Each point leads to the next one.

First, beware of the ambiguities of `is real', `exists' and %%@
similar
words. Detenserism is not just an insistence that for example, we
should use `is real' as short for `has existed or presently
exists or will exist'. And presentism is not just a
table-thumping insistence on using `exists', `is real' etc. for
`presently exists'. Rather, each doctrine assumes that some
distinction between real and unreal, in intension though not of
course in extension, is common ground to the parties to the
debate; or at least that it is common ground, as applied to
material things, events etc.---I here set aside mathematical and
other abstract objects.  Then detenserism says, with `real' (or
`exists' etc.) as applied to material items: all past, present %%@
and
future things etc. are real.  And presentism says, with the same
sense of `real' (or `exists' etc.): only present things etc.
exist.\footnote{Again I say `with the same sense of `real'', for
simplicity: it secures a direct contradiction between detenserism
and presentism. But of course different authors can and do assume
different distinctions between real and unreal; with the result
that---even if their distinctions are precise---the contradiction
between one man's detenserism and another's presentism can be
much less obvious. Indeed, their choice of distinctions might, at
a pinch, make their positions compatible.}  

Second, the debates obviously connect in various ways with those %%@
about
modality. The principal connection is via using modality to gloss %%@
the
real/unreal distinction. Thus `unreal' is often glossed as %%@
`merely
possible'. Tensers (i.e. opponents of detenserism) typically say %%@
that
the future, and maybe the past, is not actual, but merely %%@
possible.
And similarly presentists say (in terms of things, for %%@
simplicity):
Abraham Lincoln and Sherlock Holmes are on a par; so are my %%@
first-born
grandchild (supposing there is one---it is hard to know, and, not
knowing, hard to name him or her), and Darth Vader (supposed %%@
fictional,
as intended!).

This connection with modality means that the debates have been
invigorated by recent developments in modal metaphysics. In
particular, Lewis' bold advocacy of the equal reality of all
possible worlds ([1986]) gave a clear modal analogue of
detenserism; and similarly made the contrasting actualist view an
analogue of presentism. Not that these analogies made everything
cut and dried. In particular, as Lewis himself brought out: (i)
one should not just identify `being real' with `being concrete',
since the concrete-vs.-abstract distinction is itself in bad %%@
shape
([1986], Section 1.7); (ii) one cannot expect
the debates about the identity of items, through time and across
possible worlds, to be strictly parallel---not least because here
the distinctions between things, events, states of affairs etc.
come to the fore ([1986], Chapter 4).

My third point follows on from the first two. In short, it is %%@
that we
should not assimilate detenserism and presentism to various rival
semantic proposals. There is a temptation to do so; (indeed, I %%@
think
the literature of the 1950s to 1970s was wont to do so). Thus it %%@
is
tempting to say that detenserism goes with a simple bivalent %%@
semantics
of temporal discourse.  Detenserism seems to go with a semantics %%@
that,
prescinding entirely from all the complexities of natural %%@
language,
uses either (i) a single domain of quantification containing all
objects that ever exist, or (ii) a linear order of domains, each
containing the objects that exist at a single time, so that the
quantifier represents present-tensed `exists'. (Here `object' %%@
covers
things etc.) In either case, `now' and other temporal indexicals %%@
get a
straightfoward time-dependent reference. (For example: If times %%@
are
treated as objects in the domain, then `now' can be assigned a %%@
time as
reference.)  Correspondingly, tenserism and presentism seem to go %%@
with
more complex semantic proposals: say with using three %%@
truth-values, or
a branching future; or both of these.

But we should beware of the gap between semantics and %%@
metaphysics:
each discipline is and should be beholden to considerations,
substantive and methodological, that the other ignores. In the %%@
present
context, not only might linguists have reasons for or against %%@
these
semantic proposals, which ride free of metaphysics; also, the
proposals do not straightforwardly express the metaphysical %%@
positions,
just because formal semantics is not concerned with what is %%@
`real'.
Thus the use of a single big domain of quantification, as on the %%@
first
proposal, is not implied by all its members being real; so the
detenser may well endorse one of the more complex semantics. And %%@
the
tenser will note that even these proposals do not capture her %%@
metaphysical
thesis about reality.  In particular, any such semantics requires
`now' and other temporal indexicals to be treated just as they %%@
were in
the simple bivalent semantics.  It is part and parcel of doing
semantics---whether with two truth-values or more, whether with
branching or not---that such indexicals get a straightfoward
time-dependent reference.  So the `movement of the now', which %%@
for the
tenser and presentist is the crucial fact about time, is %%@
represented
only by the semantics' use of a family of interpretations, %%@
related to
each other by `sliding along' the reference assigned to `now'
etc.---exactly as in the simple semantics favoured by the
detenser!\footnote{I believe this point is not affected by the
complexities of allowing for relativization of truth-value to
circumstances of assessment, as well as circumstances of %%@
utterance;
but I cannot argue the point here.}

\subsubsection{Spontaneity} The third meaning of `time is unreal' %%@
is much less familiar in philosophy: I will call it
`Spontaneity'. For philosophers of physics, and for Barbour %%@
himself,
the most influential formulation is that of Bell ([1981]). But %%@
his
formulation is combined with a discussion of Everettian
interpretations of quantum theory (cf. Section 3.2); and %%@
Spontaneity
makes just as good (maybe better!) sense in the context of %%@
classical
physics, and even metaphysics. That is one reason why it is worth
stating Spontaneity, before we tangle with quantum theory. A more
important reason is that discussing Spontaneity will enable me to %%@
state
the essentials of Barbour's own position. Besides, some of the
comments below about Spontaneity carry over directly to
Barbour's position.

Spontaneity presupposes the idea of a set of many possible
courses of history, where each course of history is a `block
universe' in the sense of Section 3.1.1. But Spontaneity then %%@
proposes that unbeknownst to us, the actual
history jumps between disparate instantaneous states.

To explain this, suppose we are given, in metaphysics or in %%@
physical
theory, a set of possible courses of history. We naturally think %%@
of
one of these as `real', `actual', `realized' or `occupied'; %%@
(setting
aside now Section 3.1.1's issues about temporal becoming!).
And---especially in physics, if not metaphysics---we think of %%@
these
possible histories, including the actual one, as continuous in %%@
time.
That is, we think of a possible history as a sequence of %%@
instantaneous
states of the world (in metaphysics) or of the system (in %%@
physics);
and we think of the set of all possible instantaneous states as %%@
having
a topology, or some similar `nearness-structure', so that it %%@
makes
sense to talk of states being close to each other.  And because, %%@
as we
look about us, we seem to see the state of the world changing
continuously, not in discrete jumps, we naturally think that the
possible histories should be not merely sequences of %%@
instantaneous
states, but continuous curves in the (topological or similar) %%@
space of
such states. So we think of a
collection of curves, each curve representing a possible course %%@
of
history; and we think of one such curve as real, as actual---just %%@
as
in Section 2.\footnote{Two qualifications. (1) I have %%@
deliberately
avoided being precise about technical matters such as the %%@
topology of
states, and whether to require some kind of smoothness
(differentiability) as well as continuity. For they differ from %%@
one
theory to another; and make no difference to the discussion to %%@
follow.
(2) Agreed, not all of our experience, nor all our physical
theories, suggest that change is smooth or at least continuous.
Examples include Brownian motion and flourescing ions.  But so %%@
far as
I know, physics has very largely
succeeded in modelling change as continuous, and models allowing %%@
discontinuous change at least assign decreasing probability to %%@
`larger' changes; (at least once we set aside the instantaneous %%@
jumps that some invoke to
solve quantum theory's measurement problem!).}

Now I can state Spontaneity. It denies that the possible
histories (including the actual one) need to be continuous in
this sort of sense, and even that `larger' discontinuous changes %%@
need be less probable. It urges that the possible histories, in %%@
particular
the actual one, jump about arbitrarily in the space of
instantaneous states. This mind-bending doctrine calls for six
comments: rather a lot, but they will shed a lot of light on %%@
Barbour.

(1) At first sight, Spontaneity seems flatly incompatible with
our impression that the state of the world changes continuously.
But it might just be compatible. For the advocate of Spontaneity
will argue that our evidence for that impression---indeed, all
evidence for all empirical knowledge!---consists ultimately in
correlations between experiences, memories and records that are
defined at an instant. Thus: a present observation is not checked
against a previous prediction, but rather against a present
record of what that prediction was.  This predicament, that
epistemologically we are `locked in the present', implies that
any jumps of the type Spontaneity advocates would not be
perceived as such. Immediately after the jump, the new
instantaneous state, at which the actual history has arrived,
contains records fostering the illusion that the state in the
recent past was near (in the topology of the state-space) {\em
it}---and so not near the actual predecessor, which is now a jump
away.

Barbour does {\em not} advocate Spontaneity, though his denial of %%@
time
will be similar (and equally mind-bending). But he {\em does}
endorse the idea just mentioned, of our being epistemologically %%@
`locked in the
present'. For example, he writes:
\begin{quote}

But what is the past? Strictly, it is
never anything more than we can infer from present records. The
word `record' prejudges the issue. If we came to suspect that the
past is a conjecture, we might replace `records' by some more
neutral expression like `structures that seem to tell a
consistent story. (p. 33)
\end{quote}
\noindent John Bell formulates the idea similarly:
\begin{quote}
... we
have no access to the past. We have only our `memories' and
`records'. But these memories and records are in fact {\em
present} phenomena. The instantaneous configuration ...  can
include clusters which are markings in notebooks, or in computer
memories, or in human memories. These memories can be of the
initial conditions in experiments, among other things, and of the
results of those experiments. ([1981], p. 136; cf. also [1976a],
p. 95.)
\end{quote}

(2) But philosophers will recognize that this idea, that we are
`locked in the present', is very questionable. One obvious %%@
objection
one might offer against it is that even our most immediate mental
states have a duration; and for some states, a long duration may
even be necessary---can you feel deep grief for only a second?
This undermines Spontaneity's claim to be able to characterize
our evidence as records etc. at an instant.

 Here, I say `philosophers will recognize' and `obvious
objection', because Spontaneity is in effect a form of
scepticism: viz., scepticism about what occurred in the past.
More precisely: Spontaneity becomes such a form of scepticism,
if one defines it as saying, not that the actual past course of
events {\em was} a discontinuous sequence of states, jumping
about very differently from what we naively believe; but rather
that for all we know, the actual past could have been such a %%@
sequence.

  So an obvious strategy for replying to Spontaneity is to adapt
strategies fashioned for the more familiar case of scepticism %%@
about
the external world. And the objection above is just the analogue
against Spontaneity of the familiar objection against scepticism %%@
about
the external world: that our evidence cannot be characterized %%@
except
in terms of that world.\footnote{As it happens, I think the %%@
analogue
against Spontaneity is more plausible than the familiar %%@
objection.}

(3) This objection against the idea that we are epistemologically
`locked in the present' (and hence against Spontaneity) should be
distinguished from a different idea, which also seems to be
evidence against Spontaneity---but which is readily enough
answered. It is worth emphasising the distinction, since
unfortunately Barbour does {\em not} address the objection in
(2). But he does explicitly raise this second idea, and he gives %%@
the
ready answer.  For the idea seems to be evidence against his own
view, as well as Spontaneity.

This second idea is that in some cases the {\em content} of
perception requires continuity. (So the distinction from
(2) turns on the familiar contrast between a mental state and its
content.) The obvious cases involve visual perception of motion:
Barbour takes the example of watching a kingfisher in flight (pp.
17, 264). Such cases are indeed evidence against Spontaneity,
{\em if} the content is veridical---e.g. if the kingfisher really
has a continuous flight-path. 

But an advocate of Spontaneity can dig in her heels; (in an ugly
philosophical jargon, she can adopt an `error theory'). That is,
she can say that the contents of such perceptions just are {\em %%@
not}
veridical, except occasionally---when no `jump' occurs in the
relevant time-period. And she can explain away the compelling
appearance of continuous motion, by saying that the state to
which history jumps, when one judges, say, `the kingfisher is
flying to the right over the centre of the pond', involves the
simultaneous presence in the brain at that time of (delusive!)
`records' of configurations of the kingfisher a bit to the left %%@
of the
pond's centre. (Barbour's answer will be similar. He will also %%@
appeal to the
simultaneous presence in the brain of delusive records; but with
the difference that according to him, there is no actual past
motion, not even a jumpy one!)

(4) This comment and the next mostly concern Bell.  First, I
should emphasise that though Bell formulates Spontaneity, he does
not advocate it. In fact, he makes a wry comparison with another
now notorious case `where the presumed accuracy of a theory
required that the existence of present historical records should
not be taken to imply that any past had indeed occurred' (ibid.).
This case is the strategy of reconciling Archbishop's Ussher's
biblical calculation that the Earth was created in 4004 BC, with
the fossil evidence of a much greater age---by holding that God
\begin{quote}
would quite naturally have created a going concern. The trees
would be created with annular rings, although the corresponding
number of years had not elapsed. (ibid.: cf. also [1976a], p.
98).
\end{quote}

 But Bell {\em does} think Spontaneity is a natural way to %%@
understand
the Everettian interpretation of quantum theory (which he also %%@
does
not endorse). More precisely, he thinks an analogue of %%@
Spontaneity
that allows many actual histories, including discontinuous ones, %%@
is a
natural way to understand Everett; (details in Section 3.2). And
following Bell, Spontaneity has been discussed in that context; %%@
(a
recent example is Barrett ([1999]), pp. 122-127).  So those
sympathetic to Everett, if not the rest of us, need to carefully
consider Spontaneity's merits.

Here Barbour differs from Bell. Since Spontaneity and the
Everettian interpretation are both close to Barbour's own %%@
position,
he of course believes they are both close to the Truth.

(5) Spontaneity is  not presentism, as understood in Section %%@
3.1.1.
The difference is clear. Spontaneity is about dynamics: it
claims that the actual past course of events was a
discontinuous sequence (or could have been, for all we
know). Presentism is about reality: it claims that there is no
actual past course of events. (And similarly for the future.)

 But this difference is worth stressing since
Bell sometimes expresses himself in a way that suggests
presentism. He calls Spontaneity `a radical
solipsism---extending to the temporal dimension the replacement %%@
of
everything outside my head by my impressions, of ordinary
solipsism or positivism' ([1981], p. 136): words which better %%@
suit
presentism.

(6) Finally, Spontaneity faces a problem about relativity. It is
an aggravated form of the problem at the end of Section 2: viz.
the tension between having instantaneous configurations as the
fundamental ingredients of a `3+1' dynamics, and relativity's
allowing spacetime to be foliated in countless different ways. We
saw that this tension was allayed in those `3+1' theories (like
Barbour's various intrinsic dynamics, and general relativity) in
which there was a suitable `meshing'; i.e. in which any foliation
of a spacetime constructed from a given solution to the `3+1' %%@
theory
yields another such solution. But given Spontaneity, with its
allowance that histories jump about the state-space, what sense
if any can be made of such meshing?

With his usual acuity, Bell saw this problem, though he expressed
it very concisely:
\begin{quote}
The question of making a Lorentz invariant
theory on these lines raises interesting questions. For reality
has been identified only at an instant [I read this, not as
presentism, but as: instantaneous configurations are the
fundamental ingredients of dynamics---JNB] ... In a Lorentz
invariant theory would there be different realities corresponding
to different ways of defining the time direction in the
four-dimensional space? Or if these various realities are to be
seen as different aspects of one, and therefore correlated
somehow, is this not falling back towards the notion of
trajectory? ([1981], p. 136)
\end{quote}
\noindent Hard questions, which an advocate
of Spontaneity would have to face. But not Barbour: he sidesteps
the problem by denying that there is even one actual sequence of
states!

\subsubsection{Barbour's vision: time capsules} 
All the ingredients are to
hand: I can now let the cat out of the bag, and state the
essentials of Barbour's denial of time. (`Essentials' because in
Section 3.2, quantum theory will add some features.) In short, it %%@
is a
hybrid of Spontaneity and a strong realism about all the possible
instantaneous configurations---a realism analogous to Lewis' well %%@
known
realism about all possible worlds ([1986]).

Recall that Spontaneity has a single real course of history among
the many merely possible ones: its novelty is to hold that this
single actual history jumps about---instead of being a continuous
curve in state-space. On the other hand, Lewis' modal %%@
realism---adapted
to the state-space of some theory, rather than to Lewis' set of
all logically possible worlds---makes no heterodox claims about
dynamics. Its novelty is to hold that the possible courses of
history (each nicely continuous) are all equally real---instead
of just one curve in state-space being picked out as real.

Barbour proposes to go further than Spontaneity's denial that the
actual history is continuous. He denies that there is an actual
history (either past or future): there is {\em just} the space of %%@
all
possible instantaneous configurations of the universe. Here `all
possible configurations' does not mean all logically
possible configurations, but rather all the relative %%@
configurations of some suitable Machian theory (i.e. an RCS). %%@
Indeed, we will see later that for quantum theory, Barbour %%@
suggests that it means all configurations ascribed a non-zero %%@
amplitude by the quantum state; (cf. Section 3.2.1, especially %%@
(B)(2)).

And on the other hand, Barbour takes these configurations to be %%@
all
equally real, just as Lewis holds the various possible worlds %%@
(i.e.
possible courses of history) to be equally real. He of course %%@
concedes
that one can mathematically define sets of configurations; and in
particular continuous curves (since the RCS will presumably have %%@
a
topology), and even curves that obey some variational principle, %%@
as in
Section 2's theories. But these sets and curves are `just
mathematical': there is no actual physical history faithfully
represented by one of the sets---not even ({\em \`{a} la} %%@
Spontaneity) by
a discontinuous set.

That is Barbour's core idea. He obviously needs, as Spontaneity
does, to explain away our impression that there is history, and a
continuous one to boot. More specifically, he needs to argue that %%@
we are
epistemologically `locked in the present', and that the content
of any perception that requires temporal duration (e.g.
motion-perception) is, despite appearances, false. (Note that the %%@
second challenge is harder than that faced by the advocate of %%@
Spontaneity: she only needs to rule false contents that require %%@
continuity.)

As we saw in (1)-(3) of Section 3.1.2, Barbour can and does go %%@
part of
the way to doing that.\footnote{Though not all the way, as we %%@
noted at
the start of (3).} In particular, as regards the second %%@
issue---the
delusiveness of motion-perception---he takes (what we call!)
perception of a kingfisher flying to the right over a pond to %%@
involve
the brain containing a whole collection of (what we call!) %%@
records of
configurations of the kingfisher and the water. But not just any
collection. Not only are these configurations similar, i.e.  near %%@
each
other in some topology or metric on configurations, like those %%@
used in
intrinsic dynamics; also, they can naturally be given a linear %%@
order,
so that they correspond to points along a curve in the %%@
configuration
space; (pp. 29-30, 264-267). As we saw in Section 2, intrinsic
dynamics suggests how even a metric for this linear order can be
defined from just the structures of the configurations. And as we
noted before, the advocate of Spontaneity who `digs her heels in' %%@
has
a similar position: she says that motion-perception involves the
simultaneous presence in the brain of records that are misleading
about what occurred in the actual past---a similar position, %%@
except
that for her, `simultaneous' and `actual past' make sense!

Note that both positions are much more radical than the claim %%@
that
motion-perception involves the simultaneous, or roughly %%@
simultaneous,
presence in the brain, of records of very recent low-level %%@
perceptual
states. {\em This} claim is nowadays a commonplace of empirical
psychology, albeit a vague one: how else but by some sort of
integration or coarse-graining of several such records could
perception of motion be distinguished from perception of stasis? %%@
So
the difference between this claim and the radical metaphysical
positions, Spontaneity and Barbour's position, is clear. But it %%@
is
worth stressing. For Barbour's popular formulations (pp. 29-30,
266-267) blur the difference, so that the plausibility of the %%@
former
accrues invalidly to the latter.

So according to Barbour, our impression that there is history %%@
arises
from some configurations of the universe (including those we are %%@
part
of) having a very special structure: namely, they `contain %%@
mutually
consistent records of processes that took place in a past in
accordance with certain laws' (p. 31). More precisely, they %%@
contain
subconfigurations that falsely suggest such a past.  Barbour has %%@
a
memorable name for such configurations; he calls them `time %%@
capsules'.
So in short: a time capsule is any instantaneous configuration %%@
that
encodes the appearance of history, for example a history of %%@
previous
motion; and Barbour proposes that time capsules explain away our %%@
impression that there is history.

Barbour develops this vision in several ways. Perhaps the most %%@
important is his suggestion that in quantum physics, the quantum %%@
state gives time capsules relatively high probability (in which %%@
case his explanation of our impression of history is stronger, in %%@
that our impression is to be expected). I postpone this to the %%@
discussion of quantum physics in Section 3.2. But I can already %%@
explain two further points developing the idea of
a time capsule.  Barbour puts them somewhat metaphorically; (I %%@
think
they are clearest at pp. 302-305). But they are worth stating
precisely and generally; since together they give Barbour a kind
of coarse-grained surrogate of history, and they might also help %%@
him give an account of the direction of time.

(1) The first point is that one time capsule can `record' %%@
another.  That
is, one time capsule can contain records of another, without the %%@
other
similarly containing records of the first. Here the phrase `$C_1$
contains records of $C_2$' is of course colloquial---we are %%@
`speaking
with the vulgar'. According to Barbour, it is short for something %%@
like
`$C_1$ has sub-configurations that according to `vulgar' %%@
dynamical
laws are time-evolutes, or causal consequences, of some
sub-configurations of $C_2$'. Furthermore, $C_1$'s records of %%@
$C_2$
are often in its fine details, rather than its more obvious %%@
features.
The intuitive idea (with an actual history, and the records %%@
existing
later in time!) is familiar, both in science and everyday life. %%@
For
example, a rock from one epoch contains in its fine details a %%@
fossil
which records the structure of an organism that lived in some %%@
prior
epoch. The scene of the crime today contains in its fine details
fingerprints which record the suspect's being there yesterday.

(2)  The second point is that one time capsule can in this sense %%@
contain
myriadly many records of another, and even many records of one %%@
and the
same sub-configuration of the other. Again the intuitive idea is %%@
very
familiar. Many different fossils from one epoch (all
sub-configurations of one vast configuration) tell overlapping %%@
but
mutually consistent stories about some prior epoch. Today the
suspect's fingerprints are all over the scene of the crime, and
furthermore, his handkerchief soaked in the victim's blood is %%@
under
the desk. Furthermore, we must allow that in general, the %%@
different
records will not be wholly consistent with each other: geologists %%@
and
detectives often confront conflicting pieces of evidence---and %%@
today's
newspapers tell overlapping but not wholly consistent stories %%@
about
yesterday's events.

By putting points (1) and (2) together, Barbour can recover a
coarse-grained surrogate of history. We have already seen the %%@
main idea in his treatment of motion-perception: the
intrinsic structures of each of a set of configurations can %%@
define a
linear order on the set.  But now we are to suppose that the set %%@
of
configurations being considered is not just a relatively simple %%@
set of
(what we call!)  perceptual records of a moving object, but a %%@
vastly
complex set of time capsules, each containing many fine details.
Indeed, Barbour's vision is that we should consider %%@
configurations of
the whole universe.  So now any such linear order will be defined %%@
by
the way that the fine details of a configuration $C_1$ are %%@
records of
another $C_2$. It will not be defined just by a relatively simple
comparison of the more obvious features of the %%@
configurations---like
inter-particle distances or distances to portions of water, in %%@
the
simpler examples of intrinsic dynamics or perceiving a bird's %%@
flight
over a pond.

Furthermore, Barbour proposes that these fine details will not
prescribe a unique curve (linear order) through each %%@
configuration.
Again, the intuitive idea is familiar in everyday life; (and even %%@
in
science, apart from the special, albeit familiar, cases of %%@
physical
theories that are deterministic---such as Newtonian mechanics %%@
once we consider not only configurations but also their rates of %%@
change). All the fossils in all the rocks from one epoch do not
record every detail of life in the prior epoch. All the details %%@
of the
scene of the crime today may record who is the murderer, but do %%@
not
record every detail of the murder---did the murderer breathe an %%@
even
number of times while in the room? In general, today's fine %%@
details
only record some of the more obvious features of yesterday. 

But of course, Barbour, with his denial of history and belief in %%@
the
equal reality of all configurations, proposes to boldly %%@
extrapolate
this intuitive idea. According to him, the fact that today's fine
details do not prescribe a unique past is not just a matter of %%@
our
having lost information about the actual past---there was no such
past.

Barbour sums up these proposals, using the example of the %%@
possible
configurations of a swarm of bees. He takes as its obvious %%@
features
the overall position of the swarm and its size; in physics %%@
jargon, he
takes as the coarse-grained or collective variables, the position %%@
of
the centre of mass of the swarm, and the swarm's radius. The fine
details are given by the relative positions of the bees within %%@
the
swarm. So a possible (fine-grained) history of the swarm (if %%@
there
were such histories!) is given as usual by a curve in %%@
configuration
space: for 5,000 bees treated as point-particles, a space with
dimension 15,000. A coarse-grained history is then a projection %%@
of
such a curve onto the 4-dimensional subspace coordinatized by the %%@
four
coarse-grained variables: viz. the three components of the %%@
position of
the centre of mass, and the swarm's radius. Thus Barbour writes:
\begin{quote}
For each point along the
[coarse-grained history] there will be a corresponding cloud of %%@
points
that record the same history up to that point in different ways. %%@
There
will be a `tube' of such points in the configuration space. No
continuous `thread' joins up these points in the tube into %%@
Newtonian
histories. The points are more like sand grains that fill a glass
tube. Each grain tells its story independently of its neighbours. %%@
In
any section of the tube, the grains all tell essentially the same
story but in different ways, though some may tell it with small
variations. (p. 304-5).
\end{quote}

This completes my exposition of Barbour's vision (neglecting %%@
quantum physics). I will end with three comments. The first is a
suggestion to Barbour about how to treat the direction of time; %%@
the
second is an objection to his vision; and the third sums up so %%@
far. 

(a) In my discussion so far of Barbour's denial of time, I have %%@
not
had to mention the direction of time, i.e. the various %%@
asymmetries
between past and future. Indeed, this topic is not
central to his denial of time: in fact, he treats it briefly in %%@
terms of
entropy increase, though only in the context of proposals about
quantum physics (pp. 289, 318). But there is an approach to the
direction of time, within the philosophical literature about
causation, which emphasises the role of records---and which %%@
therefore
might be fruitful for Barbour. The starting-point of this %%@
approach is
the observation that, as in our examples of fossils and %%@
fingerprints,
an event typically has many later traces.  Admittedly, `trace' is %%@
here
a term of art, understood differently by different authors; for
example it might be understood as an event that is nomically
sufficient for the given event. The idea of the approach is then %%@
that we can define `trace' in a time-symmetric fashion (such as %%@
my example,
`event that is nomically sufficient for the given event'), and %%@
nevertheless maintain that an
event typically has many {\em more} later traces than it has %%@
earlier
ones.\footnote{Perhaps the best known statement of this %%@
past-future asymmetry
is by David Lewis, in the course of defending his counterfactual
analyis of causation under determinism; ([1979], pp. 49-50).} 

The relevance of this idea to Barbour is clear enough. If indeed %%@
there
is such a past-future asymmetry in traces (i.e. in Barbour's %%@
jargon,
records) then Barbour might be able to advert to it so as to give %%@
a
sense to the `tubes' that define his coarse-grained histories, %%@
and
even to the many possible curves within such a tube. However, I %%@
should
note two wrinkles. (i) It is not clear that any of the %%@
conventional
arguments for such an asymmetry, relying as they do on the %%@
existence
of time, will carry over to Barbour's framework, with its denial %%@
of
time. (ii) In any case, the relation of such an asymmetry to an
asymmetry of entropy (which Barbour believes important to %%@
past-future
asymmetry), is obscure; (as Lewis notes ([1979], p.51); cf. Sklar
([1993] pp. 401-404), for a more recent discussion).

(b) Barbour's vision leads to a curious problem about the %%@
explanation of our experience. Namely, why should we find %%@
ourselves with a perception of anything having happened, or %%@
indeed of anything now
happening?  That is, why should beings in an
`encapsulated instant' be endowed with such 
sophisticated but delusive experiences?  After all, what could %%@
these experiences be good for? Any sort of evolutionary %%@
explanation 
is obviously ruled out! (Thanks to Adrian Kent for this point.)

(c) To sum up, let me reiterate the point made at the start of
this Subsection: that while all of Section 2's classical %%@
theories,
whether traditional or Machian, fit perfectly well with %%@
detenserism,
{\em none} of them give any support at all to this Subsection's %%@
other
three meanings of the slogan that time is unreal. Now that we %%@
have
presented those meanings---presentism, Spontaneity and Barbour's
denial of time---I think the point is clear: all too clear to %%@
need
spelling out case by case, for each theory and each meaning of %%@
the
slogan.\footnote{Incidentally, such an exercise yields various %%@
minor
comments, though none helpful to the unreality of time.  For %%@
example,
some tensers and presentists may be disquieted at the emergence %%@
of the
temporal metric in intrinsic dynamics, i.e. at the idea that %%@
something
so fundamental as the `rate' at which `time passes' should be %%@
fixed by
the world's material contents (via the condition that $T = [E - %%@
V]$).} So accepting that classical theories, even Machian ones, %%@
do not
support Barbour's denial of time, one naturally asks: why on %%@
earth
should we believe it? As I said at the start of this Subsection,
Barbour's own reasons derive entirely from quantum physics---to %%@
which
I now turn.

\subsection{Evidence from quantum physics?}
At first sight, quantum theory seems as unpromising a place to
look for evidence of the `unreality of time' as were Section 2's
classical theories. In particular, there are two features of the
treatment of time in quantum theory that are often noted, and are
perhaps the most obvious features of the treatment---but neither
seems to support the unreality of time in any of Section 3.1's %%@
four
senses.

 The first feature is that quantum theory is traditionally
interpreted as indeterministic.\footnote{Though the idea that
quantum theory is  indeterministic is in fact controversial, it
is perhaps the most widely accepted `fact' about quantum theory,
among the general public; who confusedly suppose it to be part %%@
and
parcel of Heisenberg's uncertainty principle.} But indeterminism %%@
does not
support any of Section 3.1's senses: in particular, the detenser
can perfectly well admit many alternative possible futures, as
well as the actual one. The second feature is that quantum theory
treats time as a parameter `external' to the system, like the
time of Newtonian mechanics and special relativity---but unlike
general relativity or intrinsic dynamics. But that
no more supports the unreality of time than did the corresponding
treatments of time as `external', in some classical theories.

Agreed, first impressions can be deceptive; and here the danger
is all the greater since the interpretation of quantum theory is
notoriously controversial. But as I see it, none of the four main
approaches to interpreting quantum theory, and especially to
solving the main problem (of measurement), seem to support the
unreality of time. To spell this out a bit, I take the four main
approaches, as follows. One can aim to solve the measurement
problem in terms of physics, with or without revising the unitary
dynamics: this strategy yields respectively, dynamical collapse
models, or models that postulate extra `beables' such as the
pilot wave theory. Or one can aim for a distinctively
philosophical solution to the measurement problem, again with or
without revising the unitary dynamics: this strategy yields
respectively, some version of the Copenhagen interpretation, or
some kind of Everettian interpretation. 

Lack of space means I cannot here consider {\em seriatim} whether
each of these four approaches is somehow related to the unreality
of time in one of Section 3.1's senses. It must suffice to make
three comments.  First: there seem to be no close connections.
Second: in any case, no such connection was made (so far as I
know) until Bell ([1981]) suggested that Spontaneity was the
natural way to understand time in Everettian interpretations.
Third, and more importantly for us: no one before Barbour (so far
as I know) suggested that quantum theory, more specifically
quantum gravity, supported denying time in his sense (Section
3.1.3).

It will be clearest to think of Barbour's appeal to quantum
physics as proceeding in two main stages. The first stage
(Section 3.2.1) concerns the basic structure and interpretation %%@
of any
quantum theory. Here Barbour emphasises some orthodox work (going %%@
back
to the 1920s) on `semiclassical approximations' in quantum %%@
theory; and
combines this with a kind of Everettian interpretation. But he %%@
does
not claim that this work and his interpretation directly support %%@
his
denial of time. He only claims that they `make room for' his %%@
denial;
namely by implying that time capsules can get (relatively) high
probability, according to the quantum state.  The second stage
(Section 3.2.2) concerns the problem of time in quantum
geometrodynamics. In this second stage,
Barbour does argue for his denial of time: he thinks that the %%@
best
solution to the problem of time is to deny time, and `save the
appearances' by invoking time capsules and their high %%@
probability.

\subsubsection{Suggestions from Bell}
Barbour's first stage is inspired by Bell ([1981]). As we have
seen, Bell suggests that Spontaneity is the natural way for an
Everettian interpretation to treat time; and Barbour's
denial of time is close to Spontaneity. But there are both other
similarities, and other differences, between Bell and Barbour. So
it will be clearest to present Bell's points, pointing out
Barbour's responses as we go.

Bell is concerned with how best to develop an Everettian
interpretation of quantum theory. Not that Bell advocates such an
interpretation. In fact, Bell makes clear that his suggestions
are partly influenced by his sympathy for a rival, the pilot wave
interpretation. In a slogan, his overall idea is that one should
develop the Everettian interpretation as `the pilot wave
interpretation, but without the trajectories'.

This idea yields four suggestions about how to be a `good %%@
Everettian';
which I will letter (A) to (D). The first two are about the core %%@
ideas
of Everettian interpretations: here Bell suggests Everettians %%@
should
take a leaf from the `pilot waver's' book. Barbour will take the %%@
first
leaf but not the second, i.e.  accept (A) but not (B).  Bell's %%@
last
two suggestions are about time: as the phrase `without the
trajectories' hints, Bell here suggests disanalogies between the
Everettian and pilot wave interpretations. More specifically, %%@
Bell
introduces both Spontaneity (his third suggestion, (C)) and time %%@
capsules ((D)): ideas which as we
saw in Sections 3.1.2-3.1.3, Barbour further develops.

(A) Bell emphasises how natural it is that the pilot wave
interpretation takes position as its `preferred quantity', i.e.  %%@
as
the quantity that is always definite in value and whose extra %%@
values
answer the measurement problem's threat of macroscopic %%@
indefiniteness.
For it is above all the positions of macroscopic objects that we
intuitively want to be definite in value; in particular, %%@
measurement
outcomes are recorded in positions, e.g. of a pointer. %%@
Accordingly,
Bell suggests that an Everettian interpretation also does best to %%@
choose
position as its `preferred quantity', i.e. as the quantity in %%@
terms
of whose eigenstates one should resolve the interpretation's
postulated quantum statevector of the universe. 

Barbour in effect endorses this suggestion. This is evident %%@
enough
from Section 2's discussion of Barbour's treatment of classical
theories in terms of configurations. But note that the %%@
Bell-Barbour
agreement here need not concern just configurations for %%@
point-particle
theories, i.e. arrays of point-particle positions.  Though the %%@
pilot
wave interpretation is best known (and most developed) for the %%@
quantum
theory of a fixed number of particles, it can be extended to %%@
field
theory. Though the details vary, the common idea is that a
configurational variable, such as the value of a scalar field, is %%@
a
preferred quantity that has a definite value at every point in
space---cf. Barbour's treatment of field theories in Section %%@
2.2.2.2.
The difference is of course that the pilot wave interpretation %%@
also
postulates that these definite values evolve by a guidance
equation---and such a trajectory is of course anathema to
Barbour.\footnote{I should add that though Bell and Barbour may %%@
thus agree on preferring position or some similar configurational %%@
variable, their view is contentious. Various authors urge that %%@
avoiding macroscopic indefiniteness requires definite momenta as %%@
much as definite positions (i.e. localization in phase space not %%@
configuration space). Also Everettians nowadays appeal to the %%@
dynamical process of decoherence to select only an approximately %%@
preferred
quantity; though admittedly, this is often `close to' position, %%@
or to some similar configurational variable.}

(B) Bell's second suggestion, again arising from his sympathy
with the pilot wave interpretation, will need more discussion:
both because it is less welcome than his first, to both
Everettians in general and Barbour in particular, and because it
raises various philosophical issues. It concerns the most
familiar aspect of Everettian interpretations, viz. the `many
worlds': the idea, roughly, that the various components into
which the quantum statevector of the universe should be resolved
are `all real'. To be more precise, we should distinguish
mathematical statevectors and the physical situations they
purport to represent. So the idea is: each component, or perhaps
each component with non-zero amplitude, represents a `world'
which is just as `real', or `concrete', as the one apparent world
(including macroscopic objects and measurement outcomes) that we
see about us.

Bell suggests that the Everettian can simply drop this
idea: why not have just one real `world'---just as the pilot wave
interpretation has one actually possessed value of position (or
whatever corresponds in field theory), among the many
that are given non-zero amplitude by the quantum state? Thus he %%@
writes: 
\begin{quote}
It seems to me that this multiplication of universes is %%@
extravagant,
and serves no real purpose in the theory, and can simply be %%@
dropped
without repercussions ... Except that the wave is in %%@
configuration
space, rather than ordinary three-space, the situation is the %%@
same as
in Maxwell-Lorentz electron theory. Nobody ever felt any %%@
discomfort
because the field was supposed to exist and propagate even at %%@
points
where there was no particle. To have multiplied universes, to %%@
realize
all possible configurations of particles, would have seemed
grotesque. ([1981], pp. 133-134; cf. [1976a]. pp. 97-98)
\end{quote}

The first thing to say about Bell's suggestion is
that Barbour of course rejects it.  For recall (from
the start of Section 3.1.3) that Barbour advocates an analogue of %%@
the
`many worlds' idea: viz. he advocates the equal reality of all %%@
the
instantaneous relative configurations in some suitably Machian %%@
RCS; or at least the equal reality of those ascribed non-zero %%@
amplitude by the quantum state.
(We will see another reason for Barbour's rejection in (C) %%@
below.)

Apart from Barbour's disagreement with Bell, the main point I %%@
need to discuss  is simply that the Everettian `worlds' are {\em %%@
not} possible
worlds, in the sense used in modal metaphysics; but are rather
aspects, or `branches' of the single actual world---after all %%@
they are
determined by the actual quantum state of the universe. This
point is obvious enough: but it is important for us, because it %%@
bears
upon Bell's and Barbour's views---as I will spell out in three
comments.

 (1): I noted in Section 3.1.1 that the debate about whether past %%@
and
future were `real' had been invigorated by analogies with recent %%@
modal
metaphysics; and that one should not identify `being real' with %%@
`being
concrete', since the concrete-vs.-abstract distinction is itself %%@
in
bad shape.  {\em Mutatis mutandis}, the debate between `many %%@
worlds'
and Bell's `one world' alternative should be duly informed by
metaphysics.  In particular, one should be careful about this
distinction between possible worlds and aspects of the actual %%@
world;
and about whether the concrete-vs.-abstract distinction is in %%@
good
enough shape to bear on the debate. Though I cannot pursue these
issues, they are  relevant to the next two comments.

 (2): As I said in Section 3.1.3, Barbour's idea of the equal %%@
reality
of all the instantaneous configurations is like Lewis' modal %%@
realism.
But they differ in that for Barbour a physical theory, not modal
metaphysics, is to define the space of equally real %%@
possibilities,
i.e. the RCS.  (Which physical theory? In short, quantum
geometrodynamics--cf.  Section 3.2.2.)  But this point shows a %%@
further
disanalogy with Lewis' modal realism. For it shows that Barbour,
indeed any Everettian, needs more than just a theory to define %%@
the
`space of equally real possibilities': they also need a quantum
statevector (and some sort of `preferred basis' in which to %%@
resolve
it).\footnote{I should add that Barbour unfortunately does not
register these disanalogies, and sometimes seems to deny them. %%@
For
example, he says that the RCS contains `everything that is %%@
logically
possible' (p. 267). Maybe this is an artefact of writing a
popular book.}

(3) This comment follows on from (2). The state-dependence just %%@
noted
causes trouble for a suggestion that might be made: viz.  that %%@
the
`many worlds' idea has an advantage over Bell's `one world'
alternative, as regards explaining why the apparent (macroscopic)
world is as it is. I will briefly spell out the suggestion, and %%@
then
note the trouble---and thereby support Bell (and the pilot-waver) %%@
over
Barbour and the Everettians.

The suggestion is that with only one world, any such explanation %%@
must
eventually resort to one or more unexplained `brute' facts, often
facts about what the physical laws and/or initial conditions of %%@
the
universe are; and since such facts seems arbitrary, the %%@
explanation is
ultimately unsatisfactory. On the other hand, with many worlds, %%@
there
is such a fact, or more likely a group of them, for each of the
equally real worlds; and as a consequence (says this suggestion), %%@
each
such fact, or group of facts, is {\em not} arbitrary, and the
corresponding explanation is satisfactory.

  This suggestion might be supported by two analogies (perhaps in
combination).  The first is with indexicality: the many worlds %%@
idea is
supposed to make the fact that many propositions true of the %%@
apparent
world (e.g.  `the pointer reads ``1''') are false in other %%@
worlds, as
straightforward as the fact that indexical propositions (e.g. %%@
`Barbour
is here at noon, 21 June 2000') are not true at all contexts of
utterance (even all those in the apparent world!). The second %%@
analogy
is with modal metaphysics: Lewis' modal realism has been alleged %%@
to
have a parallel advantage over actualist alternatives, that it %%@
can
explain satisfactorily (as simply indexical) what actualists must
treat as arbitrary, and so as unexplained. (But note that Lewis
himself rejects the allegation: [1986], pp. 128-133.)

As it happens, I reject this suggestion, primarily because of
disagreements about what is involved in explanation. But I will %%@
not
enter into details; (cf. my [1995], 139-142, 151-154). Here I %%@
only
need to point out that the suggestion faces trouble if---as is %%@
usual
for Everettian interpretations---the set of worlds is specified %%@
by the
quantum state of the universe (say, as the worlds ascribed %%@
non-zero
amplitude), and this state is a matter of happenstance, rather %%@
than
being somehow picked out as unique. For in that case, `brute'
unexplained facts of just the kind that the suggestion wants to %%@
avoid
will reappear at the `next level' of explanation. That is to say:
although there will not be unexplained facts at the `first %%@
level',
i.e. about the apparent world being thus and so, as against some %%@
other
way (which also enjoys non-zero amplitude), there {\em will} be %%@
such
facts about what the quantum state is.\footnote{But Barbour might %%@
at a
pinch avoid this trouble. That is, he might avoid the undermining %%@
just
mentioned. For as we will see, he conjectures that the quantum %%@
state
of the universe is very special, in that it assigns high %%@
amplitude to
time capsules---and he might just take it to be uniquely picked %%@
out by
this condition.}

(C) We need not linger very long on Bell's third suggestion: that
Everettians should adopt Spontaneity. We have already covered its
essentials, and Barbour's attitude to it as `close to the Truth', %%@
in
Section 3.1.2. But the context of quantum theory prompts two %%@
further
remarks. First: discussions of Everettian interpretations often
(rightly) point out that the interpretation needs to lay down, %%@
not
only a probability distribution over the various alternatives at %%@
each
time (given by the squared amplitudes of the resolution of the
statevector at that time, in the preferred basis), but also %%@
transition
probabilities (i.e. conditional probabilities between %%@
alternatives at
different times). In such discussions, Bell's suggestion to the
Everettian about how to treat time is often taken to be, not
Spontaneity in the qualitative sense used in Section 3.1.2
(`history jumps about; or for all we know, it does'); but rather %%@
the
quantitative suggestion that `history jumps about randomly', i.e.
there are no correlations between alternatives at different %%@
times, so
that probabilities of conjunctions are products of the %%@
probabilities
of the conjuncts.

The second remark is conceptual, and concerns Barbour's denial of
time; and it relates as much to Bell's second suggestion, (B), as %%@
to
(C). It is that if Barbour is to deny time as he wishes to (cf.
Section 3.1.3), he {\em cannot} adopt Bell's `one world' %%@
suggestion
(B).  For if one adopts (B), then there is indeed an actual %%@
history (a
unique `real' trajectory through the configuration space), albeit %%@
one
that might jump about in accordance with Spontaneity. So in order %%@
to
deny time, Barbour needs to believe in the `equal reality' of %%@
enough
of his configurations to prevent such a unique real trajectory, %%@
even a
jumpy one.

(D) Bell's fourth suggestion is crucial for Barbour. Bell reviews %%@
a
standard example from quantum mechanics; his aim is primarily to
illustrate the measurement problem, but also to introduce the %%@
topic of
records for the sake of his discussion of Everettian %%@
interpretations
and Spontaneity. This example suggests to Barbour the idea that %%@
in
some cases, the quantum state assigns relatively high amplitude %%@
to
configurations that encode records of the past: i.e. to what he %%@
calls
`time capsules'. So again, there are differences, as well as %%@
agreements,
between Bell and Barbour; and it will be clearest to first %%@
outline
Bell's discussion, and then Barbour's response.

Bell's example is the analysis by Mott and Heisenberg (in %%@
1929-30) of
the formation of tracks in a cloud-chamber when a decaying %%@
nucleus
emits an $\alpha$-particle which then ionizes atoms in the %%@
chamber.
Bell's main aim in presenting this example is to exhibit what he %%@
calls
the `shifty split': how orthodox quantum theory leaves open where %%@
one
should draw the boundary between the quantum system and the %%@
classical
background---in particular, where one should apply the projection
postulate. In this example, the simplest approach is to consider %%@
only
the $\alpha$-particle as a quantum system, and treat the atoms as
classical. On this approach, the $\alpha$-particle's %%@
wave-function
would at first be a spherically symmetric wave travelling out %%@
from the
decayed nucleus. The first ionization would involve a collapse of %%@
the
wave packet, corresponding to an approximate position %%@
measurement; the
resulting wave function then propagating like a `jet', with a %%@
small
angular dispersion, along the line from the nucleus to the %%@
ionized
atom. The shape of this jet means that the second ionization %%@
would
probably be approximately colinear with the nucleus and first %%@
atom. It
would involve a second collapse of the wave packet, producing a %%@
second
jet, probably nearly parallel to the first and therefore making %%@
for a
third ionization probably nearly colinear with the first and %%@
second.
So even this simple approach gives an answer to the initial %%@
puzzle
(posed by Mott himself), why a spherically spreading wave leads %%@
to
colinear ionizations, i.e. straight tracks.\footnote{In fact, as %%@
the
$\alpha$-particle loses energy, the angular dispersion of the %%@
jets
increases, and so the ionizations tend to be less exactly %%@
colinear.}

 On the other hand, one could also treat the atoms quantum
mechanically, postponing the application of the projection %%@
postulate
until, say, the formation of a water droplet (or even later, such %%@
as
the taking of a photograph). The simplest such approach takes the
atoms as fixed and non-interacting, but with two energy levels: a %%@
ground
state and an excited state.  On this second approach, an initial
multiple-product state of the $\alpha$-particle and many atoms in
their ground states evolves into a complicated entangled state
correlating different components of the initial spherical
$\alpha$-particle wave-function with excited states of different %%@
atoms
(corresponding spatially to the $\alpha$-particle wave-function's
components). On this approach there is of course no single first
ionization; but instead a superposition of many possible first
ionizations, which then evolves with each component developing
correlated (approximately colinear) ionizations: first a second, %%@
then
a third etc. Indeed, on this approach there are no ionizations at %%@
all,
until the projection postulate is applied, say at the formation %%@
of a
water droplet. At least this is so, unless one adopts some
`no-collapse' solution to the measurement problem, such as an
Everettian or pilot-wave interpretation.

Bell points out that in this example, as in others, there are %%@
various
different choices of the boundary between quantum and classical %%@
that
make no difference to practical predictions; though of course not %%@
{\em
every} such choice gives practically correct predictions---as %%@
Bell
says, `the first kind of treatment would be manifestly absurd if %%@
we
were concerned with an $\alpha$-particle incident on two atoms %%@
forming
a single molecule' ([1981], p. 123). The reason for this %%@
agreement is
essentially the ubiquity and efficiency of the decoherence %%@
process.
But Bell of course goes on to urge that the ambiguity of this
boundary, the `shifty split', is not satisfactory in principle; %%@
and
therefore to discuss the Everettian and pilot-wave %%@
interpretations.

Barbour's interest in the Mott-Heisenberg analysis is different %%@
from
Bell's. He does not use it to motivate the Everettian's denial of %%@
the
shifty split. Rather he sees it as a promising toy-model for the
creation of time capsules, which are so central to his vision. %%@
Indeed,
he goes so far as to say that it `is more or less the %%@
interpretation
of quantum mechanics' (p. 284). To set the scene for Barbour's
argument about quantum gravity in Section 3.2.2, I need to
mention here two points about Barbour's discussion. Though these
points lead to technicalities (which we can avoid), the main %%@
ideas are
non-technical; and the first shows up a significant lacuna in %%@
Barbour's work.

(1) The first point is the remarkable fact that the
Mott-Heisenberg analysis uses the time-{\em independent}
Schr\"{o}dinger equation, which governs eigenstates, $\psi_E$ %%@
say, of
energy that are independent of time; so in position %%@
representation
they are written $\psi_E({\bf r})$ not $\psi_E({\bf r},t)$. That %%@
is to
say, Mott and Heisenberg find a time-independent wave-function
$\psi({\bf r})$ that assigns (relatively) high amplitude to %%@
colinear
ionizations; i.e. to many straight tracks radiating out from the
decaying nucleus. There is no real conflict between their use of %%@
the
time-independent equation and Bell's `narrative' account of the
creation of records, which I summarized above. At least this is %%@
so, as
regards Bell's second approach where the successive collapses of %%@
the
wave packet are not officially countenanced. For once we set %%@
aside the
measurement problem, the issue whether there is a conflict is the
technical one whether the Mott problem satisfies the conditions %%@
for
quantum mechanics' time-independent scattering theory to agree %%@
with
its (more general) time-dependent scattering theory; and in %%@
effect it
does (p. 309).

That a time-independent wave-equation can encode (through its %%@
various
solutions) varied and intricate spatial structure is no great
surprise. As Barbour emphasises, the various intricate structures %%@
of
energy eigenstates in atomic and molecular physics are determined %%@
just
by the time-independent Schr\"{o}dinger equation and the %%@
potentials
involved (which are themselves determined by the structure of the
configuration space). But for Barbour, the use of this equation %%@
is
significant for two reasons. First, he points out that %%@
quantization of
the intrinsic dynamics of point particles leads to this equation %%@
(pp.
231-2, 237, 241). Second, Barbour will later draw an analogy %%@
between
the time-independent Schr\"{o}dinger equation and the fundamental
equation of quantum geometrodynamics; for that equation is also
apparently time-independent (cf.  Section 3.2.2).

These points show up a lacuna, and perhaps an objection, for %%@
Barbour.  His work focusses on classical physical theories with %%@
matter treated as point-particles or as fields, and their %%@
quantizations in terms of configurations (i.e. wave-functions on %%@
configuration space). He says next to nothing about our best %%@
theories of matter, viz. quantum field theories. To be sure, %%@
quantum field theories can be described in terms of %%@
wave-functions on the space of all possible field configurations, %%@
and this is presumably how Barbour would like to treat them. But %%@
if so, the fact that in quantum field theory the field operator %%@
does not commute with particle-number and similar particle-like %%@
operators means that Barbour cannot expect any simple relation to %%@
the point-particle theories with which his Machian proposals, %%@
both classical and quantum, began. In particular, this might %%@
undermine these proposals' heuristic value. (Thanks to David %%@
Wallace for this point.) 

(2) The second point Barbour stresses is that the creation of %%@
records (or, in
language that is time-independent, and neutral about the collapse %%@
of
the wave packet: the assignment of high amplitude to time %%@
capsules)
requires special conditions. The most obvious one is the ordered
nature (low entropy) of the initial state: i.e. the
$\alpha$-particle's initial spherically symmetric wave-function %%@
and
the surrounding atoms being initially in their ground states. %%@
Barbour
also stresses how Mott consistently postulates outgoing, rather %%@
than
incoming waves, to represent the result of scattering. That is of
course physically reasonable: the opposite would seem as perverse %%@
as
postulating advanced solutions in electromagnetic theory---but it %%@
is
not strictly derived from the assumptions of the problem; (pp.
288-289, 310). In Section 3.2.2, I will again touch on the %%@
question how
special these conditions are. For Barbour will want the creation %%@
of
records by a Mott-like mechanism, to be generic rather than
exceptional in quantum cosmology.\footnote{Barbour also %%@
emphasises a
third point about which I disagree. He says ([1994a], p. 2890) %%@
that
unlike the usual accounts of the emergence of records using
decoherence, he takes the records to reside not in classical, but %%@
in
quantum variables---for Mott scattering, in the electrons of the
excited atoms. But surely this is a false contrast. All agree %%@
that the
micro-constituents that `seed' records are quantum in nature, but %%@
are
decohered rapidly by their environment, be it only the microwave
background; thus in Mott scattering, the atoms' coupling to their
environment, and in particular to each other, is crucial to the
formation of water droplets.}

To sum up this Section: Barbour in effect agrees with Bell's
suggestions (A) and (D), regards (C) as close to his own denial %%@
of
time---but is forced by that denial to reject Bell's (B).

\subsubsection{Solving the problem of time?}
Finally, I turn to what at the start of this Section I called %%@
Barbour's `second stage': in which he positively
argues for his denial of time (rather than just making room for %%@
it). The idea
is that denying time is the best solution to the problem of
time in quantum geometrodynamics (i.e. the approach to quantum
gravity which Barbour favours). The illusion of time is to be
explained by the quantum state of the universe assigning high
probability to time capsules.

Here I propose to cut short a story which, though fascinating, is %%@
not
only long, but also complicated and controversial. There is no %%@
space;
and besides it is told, at about the level of this paper,
elsewhere. It must suffice to say the following. Though it is
clear that quantum theory and general relativity conflict with %%@
one
another, it is very controversial how best to reconcile them. %%@
There
are several disparate motivations: that one should somehow avoid %%@
the
singularities of general relativity, or unify gravity with the %%@
other
forces, or solve the measurement problem, or avoid postulating a
spacetime continuum---to name but four! Though one can %%@
consistently
endorse several of these (e.g. all these four), in practice %%@
different
motivations prompt very different research programmes. And the
situation is not helped by the dire lack of data: the %%@
characteristic
length at which quantum gravity effects are expected to be %%@
important
(the Planck length) is as many orders of magnitude smaller than %%@
the
proton, as the proton is smaller than the Earth! 

 In any case, one strategy is to quantize general relativity. In
effect, one tries to follow in the footsteps of the quantization %%@
of
the other successful classical theory of a fundamental
force---classical electromagnetic theory. But the situation for
gravity is much more complex than for electromagnetism.  One 
first writes general relativity in the kind of `3+1' %%@
(`Hamiltonian' or `canonical') form discussed in Paragraph %%@
2.2.2.2; with
a view to then applying established methods for quantizing a %%@
classical
Hamiltonian theory.  As noted in Paragraph 2.2.2.2, writing %%@
general relativity in this 3+1 form is restrictive, since it %%@
forbids topology change. But in any case, if one adopts this %%@
approach, one naturally expects, on analogy with the way
that quantizing the classical mechanics of point-particles yields %%@
wave
mechanics, that one will get a theory which in Schr\"{o}dinger %%@
picture
has a wave-function, with a 3-geometry $h$ say and perhaps matter
fields $\phi$ as its argument: $\psi[h,\phi]$.  Hence the name
`quantum geometrodynamics'.  (The square brackets are used to %%@
reflect
the fact that the arguments $h$ and $\phi$ are themselves %%@
functions.)

At first sight, this approach succeeds: applying the quantization
methods in an informal, heuristic way, one gets equations for a %%@
$\psi$
whose arguments lie in the infinite-dimensional space of the $h$ %%@
and
$\phi$ fields. But there are horrendous problems about giving %%@
these
equations (and associated measures, inner products etc.) a proper
mathematical meaning. But suppose we set aside these technical
obstacles: still, there is a more conceptual problem, about time.

It turns out that when one writes general relativity in 3+1 form,
there are more variables in the formalism than there are physical
degrees of freedom; (we saw this implicitly in Paragraph %%@
2.2.2.2's
discussion of quotienting by the action of spatial
diffeomorphisms).  These extra variables mean that there are
constraints, i.e. equations that relate some or all of the %%@
variables
to one another. These are the momentum and Hamiltonian %%@
constraints mentioned in Section 1; they roughly correspond, %%@
respectively, to the action of spatial diffeomorphisms, and to %%@
the time-evolution. And as discussed in Paragraph 2.2.2.2, %%@
Barbour provides a Machian analysis of them.  

In the 1950s, Dirac and others developed a method for
quantizing a Hamiltonian theory with constraints: a classical
constraint $C = 0$ becomes a requirement that the quantum state %%@
is
annihilated by a corresponding operator: ${\hat C}(\psi) = 0$.  %%@
When in
the 1960s, Wheeler, DeWitt and others applied this method to %%@
general
relativity, it turned out that---{\em modulo} the technical %%@
obstacles
just mentioned---the Hamiltonian constraint, representing the %%@
time-evolution of classical general relativity,
did not become a Schr\"{o}dinger time-dependent equation of the
familiar form ${\hat H}\psi = i\hbar\/d\psi/dt$, by which the %%@
quantum
mechanical Hamiltonian ${\hat H}$ usually governs the %%@
time-development
of $\psi$. Instead, one got a
constraint equation of the form ${\hat H}(\psi) = 0$. This is the
famous Wheeler-DeWitt equation.  Apparently, it no more contains %%@
a
time-variable, than does the time-independent Schr\"{o}dinger %%@
equation
(with energy eigenvalue zero).

Since the 1960s, there has been a lot of work, trying to somehow
recover time from the Wheeler-DeWitt equation and its associated
`frozen formalism', or from related formalisms for quantized %%@
general
relativity. (Of course this work goes hand in hand with trying to
surmount the technical obstacles.) There have been three main
strategies.  (1): One tries to eliminate the extra variables %%@
(called
`solving the constraints') before quantisation, so as to identify %%@
a
time variable; so that one can then quantise, without using %%@
Dirac's
special method of constrained quantisation. Or (2): one applies
Dirac's method and so endorses the Wheeler-DeWitt equation, but %%@
tries
to identify time as a function of the variables that appear in %%@
it. Or
(3): one abandons the idea of a state-independent notion of time:
rather, time is to be an approximate concept associated with some %%@
kind
of semiclassical solution to the Wheeler-DeWitt equation. It is %%@
this
last strategy that is relevant to Barbour.\footnote{This strategy %%@
has
been studied intensively since the mid
1980s. That only this strategy is related to Barbour is hardly
surprising, at the level of slogans: since Barbour cuts the %%@
Gordian
knot of the problem of time, by denying time, he will surely %%@
reject
general strategies, like (1) and (2), for recovering it. For his
rejection, cf. pp. 244-248; for recent philosophical reviews of %%@
all
three strategies, cf. Kuchar [1999], Butterfield and Isham %%@
[1999], Belot and Earman [2001].}

Barbour's own proposal is a version of strategy (3). For he %%@
proposes
that the Mott-Heisenberg analysis is a `prototype' for the %%@
solution to
the problem of time in quantum geometrodynamics; and this %%@
analysis
uses some ideas, in particular a semiclassical state, that are %%@
central
to this strategy. But I should emphasise that he is a heterodox
follower of strategy (3). In particular, his denial of time leads %%@
him
to say that the gravitational field has three degrees of freedom,
while the conventional verdict is two (pp. 243-246); and he also
criticizes (3)'s treatment of the direction of time (pp.  %%@
257-264).

More precisely, Barbour makes two conjectures, corresponding to %%@
his
points (1) and (2) at the end of Section 3.2.1. (1'): He %%@
conjectures
that just as in elementary wave mechanics, the structure of
configuration space (and the interaction potentials defined on %%@
it),
together with the time-independent Schr\"{o}dinger equation, %%@
determine
intricately structured energy eigenstates; so also in quantum
geometrodynamics, the structure of the configuration space (which %%@
will
be a product of Paragraph 2.2.2.2's superspace and the %%@
configuration
space of matter fields), together with the Wheeler-DeWitt %%@
equation,
determine states $\psi[h,\phi]$ that are peaked on time capsules. 

(2'): He conjectures, more specifically, that the kind of state %%@
thus
determined will be a generalization of the wave-mechanical states
written down by Mott and Heisenberg. As we saw, those states are
special, in two ways: a special initial state (in fact a %%@
semiclassical
state with low entropy) is chosen, and for each scattering, only
outgoing waves are considered. Nevertheless Barbour hopes that in
quantum geometrodynamics, the creation of records (i.e.  peaking %%@
on
time-capsules) by a similar mechanism, will be generic rather %%@
than
exceptional.

In his last Chapter, he discusses these conjectures, while
admitting (p. 308, 320) that he has no hard and fast arguments, %%@
let
alone proofs. As I read him (and the corresponding discussion in
[1994a], pp. 2891-2895), he admits to having no inklings why the
special chosen state should be favoured.\footnote{Of course, %%@
various
authors (such as Hawking, Hartle and Vilenkin) have tried to give %%@
a
theoretical motivation for one `wave function of the universe' %%@
$\psi$,
rather than another; and have studied semiclassical %%@
approximations to
their heuristic formulas for $\psi$. Barbour mentions this but %%@
does
not go into it (p.312).} But he thinks the extreme complexity and
asymmetry of the configuration space might favour Mott's second %%@
kind
of `specialness': the use of outgoing waves, or their %%@
generalizations.
Here his idea is that just as in wave mechanics, the %%@
wave-function can
be significantly constrained by being required to be regular at a
boundary (e.g. being zero at spatial infinity), so in quantum
geometrodynamics the requirement that the wave-function be %%@
suitably
regular at the point or points (called `Alpha') of configuration %%@
space
that represent all space and matter contracted to a point will
prohibit states with waves `ingoing' to Alpha.

Clearly, there are a great many issues here that one could %%@
pursue; and
I must conclude.  I hope that Section 2 brought out the %%@
foundational
interest of Barbour's work on Machian themes in classical %%@
physics, and
that Section 3.1 brought out that his denial of time shows %%@
striking
intellectual imagination. Here, I shall make no bones about my %%@
main
criticism: that Barbour does not offer anything like enough %%@
evidence
for these last conjectures---though fortunately, other authors %%@
have just recently offered some.

Agreed, quantum gravity is very controversial: recall Section 1's
image of orienteering in a blizzard. But suppose we give Barbour %%@
the
chain of technical and conceptual assumptions he wants: (i) that
quantum geometrodynamics is the way to do quantum %%@
gravity---somehow
its ferocious technical obstacles can be overcome; (ii) that %%@
strategy
(3) for recovering the notion of time, is right; (iii) that %%@
Barbour's
version of strategy (3) is right, both technically (so that e.g. %%@
the
gravitational field has three degrees of freedom) and %%@
conceptually (so
that e.g. we need not construct histories as curves through
configuration space, but can rest content with coarse-grained
`tubes'). Still we need to be given an argument why some %%@
solutions
of the Wheeler-DeWitt equation should give high relative %%@
probability
to time capsules.

Agreed, physics is hard: it would be far too much to demand a %%@
general
argument applying to a realistic infinite-dimensional %%@
configuration
space (superspace). But we can reasonably ask for some kind of
toy-model of Mott scattering in quantum geometrodynamics, perhaps
using one of the much-studied finite dimensional %%@
mini-superspaces. Barbour has not given us such a model. But %%@
without a model, one can only conclude that quantum gravity gives %%@
no reason to believe Barbour's denial of time. On the other hand, %%@
some physicists influenced by Barbour have recently developed %%@
such models (Castagnino and Laura [2000], Halliwell [2000]). I %%@
cannot enter details: suffice it to say that (speaking in the %%@
temporal vernacular), Barbour can draw some hope from this recent %%@
work!

\noindent {\em Acknowledgements} For comments on a previous %%@
version, I am very grateful to: Julian Barbour, Harvey Brown, %%@
Craig Callender, Adrian Kent, Lee Smolin, Peter Morgan, Abner %%@
Shimony, Jos Uffink and David Wallace: if only I had the time and %%@
space to act on them all! And special thanks to Oliver Pooley for %%@
discussion.

\begin{center}
{\large References}
\end{center}

Barbour, J. [1982]: `Relational Concepts of Space and Time', this
journal, {\bf 33}, pp. 251-274.  Reprinted in J. Butterfield, M.
Hogarth and G. Belot eds., {\em Spacetime}, Aldershot: Dartmouth, %%@
pp.
141-164.

Barbour, J. [1989]: {\em Absolute or Relative Motion?}, volume 1,
Cambridge: Cambridge University Press.

Barbour, J. [1994]: `The timelessness of quantum gravity: I. The
evidence from the classical theory', {\em Classical and Quantum
Gravity}, {\bf 11}, pp 2853-2873.

Barbour, J. [1994a]: `The timelessness of quantum gravity: II. %%@
The
appearance of dynamics in static configurations', {\em Classical %%@
and
Quantum Gravity}, {\bf 11}, pp. 2875-2897.

Barbour, J. [1999]: {\em The End of Time: The Next
Revolution in Our Understanding of the Universe}. London:
Weidenfeld and Nicholson. 

Barbour, J. [1999a]: `The Development of Machian Themes in the
Twentieth Century', in {\em The Arguments of Time}, ed. J.
Butterfield, British Academy and Oxford University Press, pp. %%@
83-110.

Barbour, J. and Bertotti, B. [1982]: `Mach's principle and the
structure of dynamical theories', {\em Proceedings of the Royal
Society of London}, {\bf A 382}, pp. 295-306.

Barbour, J. and \'{O} Murchadha, N. [1999]: `Classical and %%@
Quantum
Gravity on Conformal Superspace', gr-qc/9911071.

Barrett, J. [1999]: {\em The Quantum Mechanics of Minds and %%@
Worlds},
Oxford: Oxford University Press.

Bell, J. [1976]: `How to Teach Special Relativity', in Bell %%@
[1987], pp. 67-80.

Bell, J. [1976a]: `The Measurement Theory of Everett and de %%@
Broglie's
Pilot Wave', in {\em Quantum Mechanics, Determinism, Causality %%@
and
Particles}, ed. M. Flato et al. Dordrecht: Reidel, pp. 11-17.
Reprinted in Bell [1987]; page references to reprint.

Bell, J. [1981]: `Quantum Mechanics for Cosmologists', in {\em %%@
Quantum
Gravity II}, eds. C. Isham, R.Penrose and D. Sciama, Oxford: %%@
Clarendon
Press, pp. 611-637.  Reprinted in Bell [1987]; page references to
reprint.

Bell, J. [1987]: {\em Speakable and Unspeakable in Quantum %%@
Mechanics},
Cambridge: Cambridge University Press.

Belot, G. [1999]: `Rehabilitating Relationism', {\em %%@
International
Studies in the Philosophy of Science}, {\bf 13}, pp. 35-52. 

Belot, G. [2000]: `Geometry and Motion', this journal,
{\bf 51} pp. 561-595.

Belot, G and Earman, J [2001]: `Presocratic Quantum Gravity', in %%@
Callender and Huggett [2001].

Brown, H. [1993]: `Correspondence, Invariance and Heuristics in %%@
the
Emergence of Special Relativity', in {\em Correspondence, %%@
Invariance
and Heuristics}, eds. S. French and H. Kamminga, Dordrecht: %%@
Kluwer
Academic, pp. 227-260. Reprinted in J. Butterfield, M.  Hogarth %%@
and G.
Belot eds., {\em Spacetime}, Aldershot: Dartmouth, pp. 205-238.

Brown, H. [1996]: `Mindful of Quantum Possibilities', this %%@
journal,
{\bf 47} pp. 189-200.

Brown, H. [1997]: `On the Role of Special Relativity in General
Relativity', {\em International Studies in the Philosophy of %%@
Science},
{\bf 11}, pp. 67-81.

Brown, H. and Pooley, O. [2001]: `The Origins of the Spacetime %%@
Metric:
Bell's `Lorentzian Pedagogy' and its Significance in General
Relativity', forthcoming in {\em Physics Meets Philosophy at the
Planck Scale}, eds. C. Callender and N. Huggett, Cambridge: %%@
Cambridge
University Press; gr-qc/9908048.

Butterfield, J. [1984]: `Seeing the Present', {\em Mind} {\bf %%@
93}, pp.
161-176. Reprinted in {\em Questions of Time and Tense}, ed. R. %%@
Le
Poidevin, Oxford: Oxford University Press, pp. 61-75.

Butterfield, J. [1995]: `Worlds, Minds and Quanta', {\em %%@
Aristotelian
Society Supplementary Volume} {\bf 69}, pp.  113-158.

Butterfield, J. and Isham, C. [1999]: `On the Emergence of Time %%@
in
Quantum Gravity', in {\em The Arguments of Time}, ed. J. %%@
Butterfield,
British Academy and Oxford University Press, pp. 111-168.

Callender, C. and Huggett, N. eds. [2001]: {\em Physics Meets
Philosophy at the Planck Scale}, Cambridge: Cambridge University
Press.

Castagnino, M. and Laura, R. [2000]: `Functional Approach to %%@
Quantum Decoherence and the Classical Final Limit: the Mott and %%@
Cosmological problems', {\em International Journal of Theoretical %%@
Physics} {\bf 39} pp. 1737-1765; gr-qc/0006012.

Earman, J. [1989]: {\em World Enough and Spacetime}, Cambridge %%@
Mass.:
MIT Press.

Einstein, A. [1949]: `Autobiographical Notes', in {\em Albert
Einstein: Philosopher-Scientist}, ed. P.A. Schilpp, La Salle, IL: %%@
Open
Court, pp. 1-96.

Forbes, G. [1987]: `Places as Possibilities of Location', {\em %%@
Nous},
{\bf 21} pp. 295-318.

Halliwell, J. [2000]: `Trajectories for the Wave Function of the %%@
Universe from a Simple Detector Model', gr-qc/0008046.

Kuchar, K. [1999]: `The Problem of Time in Quantum %%@
Geometrodynamics',
in {\em The Arguments of Time}, ed. J. Butterfield, British %%@
Academy
and Oxford University Press, pp. 169-195.

Langton, R. and Lewis, D. [1998]: `Defining `Intrinsic'', {\em
Philosophy and Phenomenological Research}, {\bf 58} pp. 333-345;
reprinted in D. Lewis [1999]: {\em Papers in Metaphysics and
Epistemology}, Cambridge: Cambridge University Press, pp. %%@
116-132.

Lewis, D [1979]: `Counterfactual Dependence and Time's Arrow', %%@
{\em
Nous}, {\bf 13} pp. 455-476. Reprinted in his {\em Philosophical
Papers}, volume II, Oxford: Oxford University Press, 1986; page %%@
reference to
reprint.

Lewis, D. [1986]: {\em On the Plurality of Worlds}, Oxford: %%@
Blackwell.

Peacocke, C. [1979]: {\em Holistic Explanation}, Oxford: Oxford
University Press.

Pooley, O. [2001]: `Relationism Rehabilitated?', this journal, %%@
{\bf ??} pp. ??.

Sklar, L. [1993]: {\em Physics and Chance}, Cambridge: Cambridge
University Press.

\end{document}